\documentclass[
 reprint,
 amsmath,amssymb,
 aps,
 pra,
 longbibliography,
]{revtex4-1}
\usepackage{graphicx}% Include figure files
\usepackage{dcolumn}% Align table columns on decimal point
\usepackage{bm}% bold math
\usepackage{svg}
\usepackage{amsmath}
\usepackage{url}
\DeclareUnicodeCharacter{2212}{-}

\begin{document}
\preprint{APS/123-QED}

\title{Simulation of single hole spin qubit in strained triangular FinFET quantum devices}

\author{Ilan Bouquet}
 \email{bouqueti@iis.ee.ethz.ch}
\author{Jiang Cao}%
\author{Mathieu Luisier}

\affiliation{%
Integrated Systems Laboratory, ETH Zurich, Gloriastrasse 35, 8092 Zurich, Switzerland\\
}%

\date{\today}

\begin{abstract}

Using an in-house Schrödinger-Poisson (SP) solver, we investigate the creation of a single hole spin qubit inside a triple-gate triangular silicon fin field-effect transistor (Si FinFET) quantum device similar to experimental structures. The gate-induced formation of the required quantum dot (QD) is monitored based on the Luttinger-Kohn $6\times6$ $k\cdot p$ method accounting for magnetic fields and strain to determine the qubit ground state. Strain arises from the inhomogeneous contraction of the different FinFET’s components when they are cooled down to cryogenic temperatures. It leads to a renormalization of the qubit’s energy levels, thus impacting both the heavy-hole (HH) and light-hole (LH) populations as well as their mixing. The dot length, band mixing, g-factor, and Larmor/Rabi frequencies of the considered device are extracted. In particular, we show that these metrics exhibit strong strain-dependent variations of their magnitude, thus underlying the importance of including realistic thermal contraction scenarios when modeling hole spin qubits.

\end{abstract}

\maketitle

\section{\label{sec:Introduction}Introduction}

Information encoding through the electron or hole spin appears as a promising approach to achieve universal quantum computing \cite{loss_quantum_1998,burkard_coupled_1999,kloeffel_prospects_2013}. Spin-based quantum computing systems rely on the implementation of one to multiple quantum dots (QD) within a single device unit. These QDs behave as individual atoms with discrete energy states at cryogenic temperatures. The already well-established silicon (Si) complementary metal-oxide-semiconductor (CMOS) technology is ideally suited to host QDs \cite{hanson_spins_2007,zwanenburg_silicon_2013,RevModPhys.95.025003}. The semiconductor industry is indeed capable of producing electronic circuits consisting of densely-packed nanoscale field-effect transistors (FETs). By carefully engineering the gate electrodes of these Si FETs, QDs can be electrostatically formed inside their channel. Such features make Si a compelling platform to realize large-scale quantum processors. Pioneer works realized on Si/SiGe planar heterostructures achieved successful implementation of single and double logical quantum gates, which build the core of quantum algorithms outperforming their classical counterparts \cite{kawakami_electrical_2014,takeda_fault-tolerant_2016,zajac_resonantly_2018,watson_programmable_2018}. Using isotopically-enriched Si in Si-MOS devices, fidelities reaching 99\% and 85\% were reported for single- and double-qubit, respectively, alongside with decoherence times up to a few tens of microseconds \cite{veldhorst_two-qubit_2015,huang_fidelity_2019,petit_universal_2020,yang_operation_2020}. Eventually, combining the quantum well design of Si/SiGe heterostructures and the isotopically enriched Si of Si-MOS devices led to remarkable improvements in the gate fidelity of one- and two-qubit gates overcoming the error correction threshold, thus demonstrating fault-tolerant quantum computing \cite{noiri_fast_2022,mills_high-fidelity_2022}. More recently, QD systems involving germanium (Ge) as hosting material have also been extensively investigated as hole spin qubit platforms \cite{scappucci_germanium_2020,sammak_shallow_2019}. Among possible configurations, Ge/SiGe planar heterostructures \cite{hendrickx_single-hole_2020,hendrickx_fast_2020,hendrickx_four-qubit_2021,hendrickx_sweet-spot_2024} and Ge/Si core/shell nanowires \cite{froning_strong_2021,kloeffel_direct_2018} appear as leading candidates owing to their increased spin-orbit interaction (SOI) that provides fast qubit driving in the nanosecond range. In addition, early experimental works on planar gallium arsenide (GaAs) also reported the formation of gate-induced QD \cite{elzerman_single-shot_2004,nowack_coherent_2007,petersson_quantum_2010}, highlighting the potential of various semiconductor platforms as building blocks of quantum architectures. 

In this work, we focus on hole spin qubits hosted within Si FinFETs. Their electromagnetic properties can be accurately described by perturbatively mixing the wave functions of the six top valence $p$-orbitals of the underlying semiconductor \cite{fang_recent_2023}. Hole states display key features that enable the engineering of long-lifetime and scalable qubits \cite{fang_recent_2023,kobayashi_engineering_2021}. For instance, $p$-orbitals, because their spatial distribution reaches its maximum away from the unit cell center, are less sensitive to hyperfine interactions with the nuclear spin bath of the hosting material \cite{prechtel_decoupling_2016}. Moreover, it is well-known that the inherently anisotropic shapes of $p$-orbitals induce a strong intrinsic spin-orbit coupling (SOC) in contrast to the weak SOC observed in the conduction bands of this material \cite{winkler_spin-orbit_2003,froning_strong_2021,kloeffel_strong_2011,wang_ultrafast_2022}. Therefore, the interaction between the particle’s spin and its orbital motion around the nucleus can be harnessed by applying an electromagnetic signal that manipulates the spin orientation. This technique is known as electric-dipole-spin-resonance (EDSR). In terms of scalability, the intrinsic SOC of hole states represents a considerable advantage over electron-based systems where an extra micro-magnet is required to artificially enhance the SOC. Also, the chip area to implement Si hole spin qubits can be minimized by compactly integrating the manipulation and read-out mechanisms within the same quantum unit. Recent works demonstrated the use of split- and double-gate architectures in $p$-type Si FinFETs to achieve this double functionality. The first gate is used to encode the information in its adjacent qubit, whereas dispersive read-out of the electrostatically coupled qubit is performed via gate reflectometry on the second electrode \cite{maurand_cmos_2016,crippa_gate-reflectometry_2019,ezzouch_dispersively_2021,camenzind_hole_2022}. Rapid dephasing of the hole spin qubit through charge noise caused by microwave modulations of the gate voltage has been identified as a disadvantage of strong intrinsic SOC \cite{cohen_quantum_2021,wang_ultrafast_2022}. However, theoretical studies revealed that this drawback can be partially alleviated by selecting appropriate crystallographic directions of the FinFET and by taking advantage of strain \cite{bosco_fully_2021,bosco_hole_2021}. Furthermore, it was reported that devices with a triangular cross section and elongated QDs lead to significant improvement of the spin qubit lifetime over those possessing a square geometry \cite{bosco_hole_2021,camenzind_hole_2022}.

All these constraints make the fabrication, design, and implementation of quantum circuits a challenging task. While experimental activities focus on the exploration of various routes to give rise to scalable quantum devices, theoretical studies are concerned with the development of physical models to predict their working principles. An accurate modeling platform is therefore key to support the on-going experiments and to help design next-generation quantum devices. To reach this goal, a versatile tool that can describe the electronic properties of QDs and capture quantum mechanical effects within complex, experimental-like geometries is indispensable. Such a tool allows to test the influence of multiple design parameters on the performance of quantum devices, e.g., shape, dimension, crystallographic orientation, gate architecture, materials composition, or field orientation with minimal overhead. Overall, in-silico investigations can reduce the number of costly fabrication cycles and pave the way for the extraction of meaningful device metrics, thus enabling to benchmark different device configurations or even qubit platforms. 

\begin{figure}
\includegraphics[width=250pt]{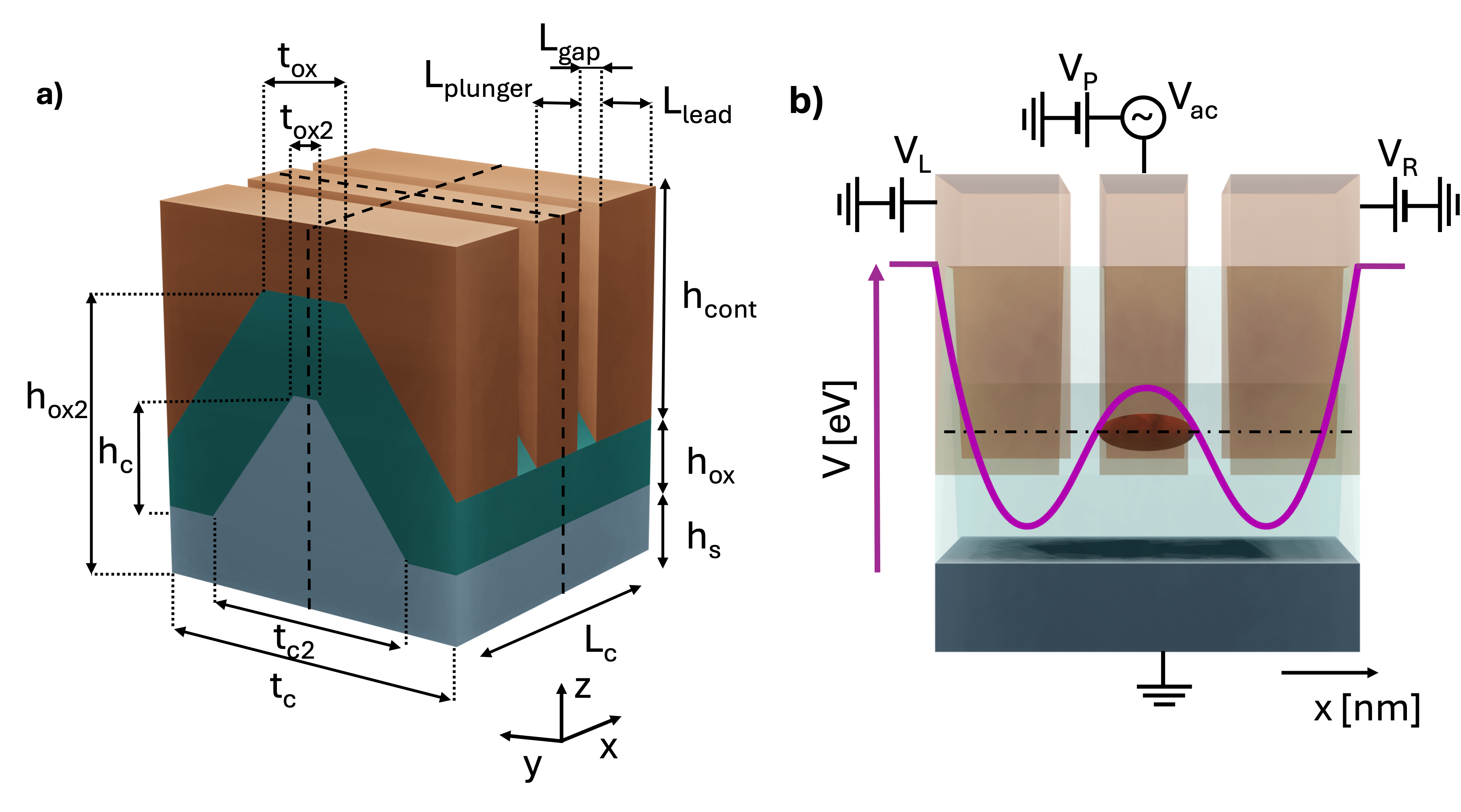}
\caption{\label{fig:device}a) Illustration of the quantum device (Si FinFET) simulated in this work. The TiN metallic gates are shown in orange, the SiO$_{2}$ layer in green, and the Si regions in blue. The structure has the following dimensions in nm: $L_{c}=50$, $L_{lead} = 15$, $L_{plunger} = 10$, $L_{gap} = 5$, $h_{s} = 8$, $h_{ox} = 4$, $h_{ox2} = 23$, $h_{cont} = 27$, $h_{c} = 15$, $t_{c} = 38$, $t_{c2} = 28$, $t_{ox} = 6$, and $t_{ox2} = 6$. b) Side view of the quantum device in a) showing the Si channel (transparent), the gate-induced QD (red ellipse), and the substrate (opaque blue). The purple curve represents the average electrostatic potential (valence band) across the transistor channel and is controlled by three metallic gates (orange): the plunger gate $V_p$ to which an AC signal $V_{ac}$ can be applied and the lead gates $V_L$ and $V_R$.}
\end{figure}

Although 3D Schrödinger-Poisson solvers implementing the 6-band kp method and including strain via the Pikus-Bir Hamiltonian have been reported before to investigate hole spin qubits in planar quantum architectures as for instance in \cite{martinez_hole_2022,abadillo-uriel_hole-spin_2023}, we believe that the consideration of cooling induced strain and its impact on the qubit figures of merit has not been studied before in FinFET and nanowire like systems. As such, the investigation of this important effect represents the main innovation of this paper. Here, we apply a quantum mechanical approach to the simulation of a triple-gate, $p$-type, triangular FinFET-like structure similar to experimental devices to reveal its functionality and assess its potential as quantum computing platform. Since the materials constituting Si FinFETs do not contract at the same rate when cooled down to cryogenic temperatures, we explore different strain configurations to model the influence of this effect on the behaviour of hole spin qubit. It is found that strain profoundly affects the nature of the qubit ground-state and changes the magnitude of the investigated device metrics. We show that, upon cooling, the choice of boundary conditions (BCs) in the linear elastic theory model leads to either thermally-induced compressive or tensile stress inside the quantum device. This in turn alters the dominant band character in the QD, e.g., HH (LH) in case of tensile (compressive) strain. By computing the principal $g$-factors and Rabi frequencies of single dot systems and comparing the results with experiments, it appears that the compression scenario is more likely to occur when the device is cooled down to cryogenic temperatures. 

The paper is organized as follows. In Section \ref{sec:Methodology}, we present the FinFET’s geometry of interest and introduce the models developed to simulate the creation of QDs in such structures, extract relevant data, and include strain. In Section \ref{sec:Results}, we discuss our results with a strong emphasis on the role played by strain on the formation of electrostatically induced QDs. Finally, we compare the device metrics extracted from our simulation to those obtained experimentally for a similar device geometry \cite{camenzind_hole_2022} before concluding in Section \ref{sec:Conclusion}.

\section{\label{sec:Methodology}Methodology}
\subsection{\label{subsec:Device}Device Structure (Si FinFET)}

The quantum device under investigation in this paper is presented in Fig.~\ref{fig:device}. It is similar to the experimental structure reported in \cite{camenzind_hole_2022} with scaled dimensions to allow for faster simulations. It consists of a FinFET with a triangular-shaped, slightly $p$-doped (10$^{14}$ cm$^{-3}$) Si channel surrounded by an SiO$_{2}$ insulating layer. The channel’s cross-section is 28 nm wide and 15 nm high. Note that its top edge is truncated and has a width of 6 nm. The SiO$_{2}$ layer is 8 nm thick and is topped by three TiN metallic gates. The central gate of length 10 nm acts as plunger gate ($V_P$) and determines the QD’s location. The 15 nm long side gates ($V_{L,R}$) are separated from the plunger gate by a gap of 5 nm and used to electrostatically define the extended reservoirs with high hole concentrations. The bottom substrate has a height of 8 nm. For simplicity, it is assumed that the qubit’s wave function cannot penetrate inside the substrate and the oxide. The device axes ($x$: channel axis, $y$ and $z$: lateral confinement) are aligned with the principal crystallographic directions, i.e., $x$ = [100], $y$ = [010], and $z$ = [001]. As compared to the experimental FinFET of \cite{camenzind_hole_2022}, the length of the source and drain extensions was reduced from 130 to 48 nm and the number of electrodes from five to three, giving rise to a single QD instead of a coupled double quantum dot (DQD). The gaps between the gates remain the same with a slightly down-scaled cross section to minimize the computational burden. Moreover, to numerically stabilize the potential away from the trapped charge, the device was extended on boths side with two, 2-nm-thick and highly p-doped (10$^{20}$cm$^{-3}$) Si regions to emulate the flat potential originating from the metallic source and drain electrodes.    

\subsection{\label{subsec:General simulation} General simulation approach}

Here, we present the simulation framework we developed to characterize the electro-magnetic properties of the single hole spin qubit induced in the FinFET structure shown in Fig.~\ref{fig:device}. The energy and corresponding wave function of the qubit ground state as well as the electrostatic potential in which it operates are determined with a home-made self-consistent Schrödinger-Poisson (SP) solver relying on the $6\times6$ Lüttinger-Kohn $k\cdot p$ method expressed on a finite-difference (FD) grid \cite{luisier_full-band_2008}. Because of the triangular shape of our quantum device we expect the electrostatic potential to exhibit a non-trivial spatial dependence that can hardly be captured by typical analytical models based, for example, on uniform electric field distributions and harmonic approximations. In contrast, conducting self-consistent Schrödinger-Poisson simulations allows one to properly describe shape and boundary effects arising from non-regular cross-sections and complex gate electrode configurations. Eventually, after convergence of the Schrödinger and Poisson equations, an accurate picture of the qubit electrostatic landscape is obtained, regardless of the device geometry or applied voltage. The $x$, $y$, and $z$ axes are discretized on a homogeneous grid with a spacing (in nm) $dx$ = 0.5 , $dy$ = 0.5 , and $dz$ = 0.5 between adjacent points. Such a resolution ensures accurate results in Si. To further minimize the computational burden, hard-wall boundary conditions are set all around the Si channel, i.e., the Schrödinger equation is restricted to the discretization points located inside the Si channel. On the other hand, Poisson's equation is solved in the full 3D domain with von Neumann BCs imposed on the external faces, except at the locations of the gate, source, and drain electrodes. The chosen Hamiltonian accurately captures the band mixing between the heavy-hole (HH), light-hole (LH), and split-off spin-orbit (SO) of the hole state. In our implementation, it is expressed in the following basis: $\{|\frac{3}{2},+\frac{3}{2}\rangle, |\frac{3}{2},+\frac{1}{2}\rangle, |\frac{3}{2},-\frac{1}{2}\rangle,|\frac{3}{2},-\frac{3}{2}\rangle,|\frac{1}{2},+\frac{1}{2}\rangle,|\frac{1}{2},-\frac{1}{2}\rangle\}$.

After convergence of the self-consistent SP simulations, an estimate of the QD $l_{dot}$ can be obtained from 
 
\begin{eqnarray}
s_{w}(r)=\sqrt{\frac{\sum_{i=1}^{n_r}w_{i}(r_{i}-\Bar{r_{w}})^{2}}{\frac{M-1}{M}\sum_{i=1}^{n_r}w_{i}}}=\frac{l_{r}}{2},
\label{eq:one}
\end{eqnarray}

$s_{w}$ represents the weighted standard deviation, $\Bar{r_{w}}$ is the weighted average position of the qubit along the direction $r=x,y,z$. The sum over $i$ includes all discretization points along one direction ($x$, $y$, or $z$), whereas $n_r$ is the total number of discretization points along this direction. Finally, $w_{i}$ is the weight (charge density) at position $r_{i}$ and $M_r$ the number of non-zero weights in direction $r$, i.e., the number of discretization points located inside the Si channel along $r$. After extraction of the qubit ground state, we calculate $\l_{dot} = (l_x+l_y+l_z)/3$ and we can decompose the wave function into basis states to obtain the relative band mixing.

The magnetic-field contribution is implemented according to Lüttinger’s method \cite{luttinger_quantum_1956}\cite{willatzen_k_2009}: 

\begin{eqnarray}
H_{z}= \mu_{B}(\boldsymbol{\sigma}-(1+3\kappa)\boldsymbol{L})\cdot\boldsymbol{B} = 2\kappa\mu_{B}\boldsymbol{J}\cdot\boldsymbol{B}.
\label{eq:two}
\end{eqnarray}

Here, $\boldsymbol{B}$ refers to the applied magnetic field and $\mu_{B}$ to the Bohr’s magneton, while $\kappa$ is the isotropic magnetic Lüttinger parameter. Finally, $\boldsymbol{\sigma}$, $\boldsymbol{L}$, and $\boldsymbol{J}$ stand for the spin (i.e., Pauli matrices), orbital, and total angular momentum operators, respectively. The effect of the magnetic field on the qubit Bloch functions is accounted for through the canonical transformation $\boldsymbol{p}\rightarrow-i\hbar\boldsymbol{\nabla}+e\boldsymbol{A}$ where $\boldsymbol{A} $ is the vector potential. In our SP simulations we use the following gauge: $\boldsymbol{A}=-(B_{z}y,0,B_{y}x−B_{x}y)$.
The Zeeman energy splitting $\Delta E$ is given by

\begin{eqnarray}
\Delta E=g^{*}\mu_{B}|\boldsymbol{B}|,
\label{eq:three}
\end{eqnarray}
where $g^{*}$ is the direction-dependent effective $g$-factor. It is linked to the symmetric Zeeman tensor $G$ through:

\begin{eqnarray}
\Delta E^{2}=\mu_{B}^{2}(\boldsymbol{B^{T}}\cdot G\cdot\boldsymbol{B}).
\label{eq:four}
\end{eqnarray}

In our approach, the SP simulations are first solved without any magnetic field and then under the influence of a small homogeneous static field. This allows to directly assess the effect of the field on the spin-orbit basis states and the energy splitting by computing the qubit’s eigenmodes and eigenergies. Following the procedure described in \cite{crippa_electrical_2018}, the $G$ tensor can then be reconstructed by extracting the $\Delta E$ arising from the applied magnetic field along the six independent spatial directions $x$, $y$, $z$, $xy$, $xz$, and $yz$. The diagonalization of $G$ into $V^{T}\tilde{G}V$ produces the square of the principal $g$-factors (eigenvalues) in the diagonal matrix $\tilde{G}$ and the principal magnetic axes (eigenvectors) in $V$ \cite{venitucci_electrical_2018}. Hence, knowing the energy splitting for a sub-set of the independent directions enables to determine the $g^{*}$-dependence along all possible directions. Consequently, substantial computational time can be saved. 

According to the method of the $g$-matrix formalism from \cite{venitucci_electrical_2018}, the Rabi frequency $f_{R}$ can be calculated by post-processing the degenerated qubit’s eigenvector pairs computed at slightly different plunger-gate potentials (i.e., at $V_{0}$ and $V_{0} \pm\delta V$) and by using the fluctuation model equation: 

\begin{eqnarray}
f_{R}=\frac{\mu_{B}BV_{ac}}{2h|g^{*}|}[\hat{g}(V_{0})\cdot \boldsymbol{b}]\times[\hat{g}'(V_{0})\cdot \boldsymbol{b}].
\label{eq:five}
\end{eqnarray}

Here, $V_{ac}$ is the amplitude of the electro-magnetic signal, $\hat{g}(V_{0})$ and $\hat{g}’(V_{0})$ are the $g$-matrix and its derivative with respect to the applied gate potential $V_{0}$. Finally, $g^{*}$ is defined as $\hat{g}^{*}=|\hat{g} \cdot b|$ where $\boldsymbol{b}$ is the unit vector directed along the magnetic field $\boldsymbol{B}$.  

\subsection{\label{subsec:Strain}Inclusion of thermally induced strain contraction}

The materials that make up the quantum device of Fig.~\ref{fig:device} (metal, oxide, semiconductor) undergo different thermal contraction rates when they are cooled down to cryogenic temperatures. Since the resulting strain might change the channel's band structure and therefore the qubit ground-state energy, it is essential to include this effect in our electro-magnetic simulations. This requires first to accurately describe the strain distribution within the FinFET and its surrounding based on thermo-mechanical models. Therefore, the quantity of interest that needs to be determined is the displacement field $\boldsymbol{u}(r)=(u_x,u_y,u_z)$ whose entries $u_x$, $u_y$, and $u_z$ contain the displacements along the $x$, $y$, and $z$ axes, respectively. From the position-dependent $\boldsymbol{u}(r)$, all entries of the strain tensor 
$\boldsymbol{\varepsilon}=\begin{bmatrix}\varepsilon_{xx}&\varepsilon_{yy}&\varepsilon_{zz}&\varepsilon_{xy}&\varepsilon_{xz}&\varepsilon_{yz} 
\end{bmatrix}^{T}$ 
can be computed at each location using:
\begin{eqnarray}
    \boldsymbol{\varepsilon}=D\boldsymbol{u},
    \label{eq:seven}
\end{eqnarray}
where $D$ is a $6\times3$ matrix containing the following entries:
\begin{eqnarray}
D=
\begin{pmatrix}
\partial_x & 0 & 0 \\
0 & \partial_y & 0 \\
0 & 0 & \partial_z \\
\partial_y/2 & \partial_x/2 & 0 \\
\partial_z/2 & 0 & \partial_x/2 \\
0 & \partial_z/2 & \partial_y/2
\end{pmatrix}.
\label{eq:sevenc}
\end{eqnarray}

The strain tensor $\boldsymbol{\varepsilon}$ is then related to the stress tensor $\boldsymbol{\sigma}=\begin{bmatrix}\sigma_{xx}&\sigma_{yy}&\sigma_{zz}&\tau_{xy}&\tau_{xz}&\tau_{yz}\end{bmatrix}^{T}$ through Hooke's law: 
\begin{eqnarray}
\boldsymbol{\sigma}=C(E,\nu)\boldsymbol{\varepsilon}.
\label{eq:sevenc}
\end{eqnarray}
The required $6\times6$ elasticity matrix $C$ depends on $E$ and $\nu$, the Young modulus and the Poisson coefficient, respectively. 

Finally, $\boldsymbol{\sigma}$ obeys the following system of linear equations
\begin{eqnarray}
    D^T\boldsymbol{\sigma}=D^T\cdot C\cdot D\boldsymbol{u}=\boldsymbol{f},
    \label{eq:sevenc2}
\end{eqnarray}
where the vector $\boldsymbol{f}=\begin{bmatrix}f_{x}&f_{y}&f_{z}\end{bmatrix}^{T}$ contains the body force terms along the $x$, $y$, and $z$ directions. In the framework of this paper, they represent thermal loads.

This mechanical problem of constrained thermal deformation induced by a change of temperature under imposed BCs has been extensively investigated. For example, a comprehensive analysis of its finite element method (FEM) formulation can be found in \cite{pepper_finite_2017,logan_first_2007}. In essence, the quantity $\boldsymbol{u}$ is expanded in terms of the tetragonal elements constituting the FEM mesh. 
The linear system in Eq.~(\ref{eq:sevenc2}) then becomes:
\begin{eqnarray}
K\boldsymbol{u}=\boldsymbol{F_{th}},
\label{eq:sevend}
\end{eqnarray}
which yields $\boldsymbol{u}$. Here, $K$ is the stiffness matrix and \boldsymbol{$F_{th}$} is the vector of thermal forces arising from the change of temperature. It is proportional to $\boldsymbol{\varepsilon_{th}}$, the vector of thermally-induced strain. In linear thermal expansion and linear elastic material theory, this vector is given by:
\begin{eqnarray}
\boldsymbol{\varepsilon_{th}}=\begin{pmatrix}
\alpha(T_{cool}-T_{rt}) \\
\alpha(T_{cool}-T_{rt}) \\
\alpha(T_{cool}-T_{rt}) \\
0 \\
0 \\
0
\end{pmatrix},
\label{eq:seveng}
\end{eqnarray}
where $\alpha$ is the temperature-independent thermal expansion coefficient, while $T_{rt}$ and $T_{cool}$ are the room (initial) and cryogenic (final) temperatures, respectively.
Note that only direct and not shear strain is generated by the temperature gradient. Finally, by imposing BCs on $\boldsymbol{u}$ and its derivative, the thermally induced displacement field can be computed and subsequently used to calculate the strain with Eq.~(\ref{eq:seven}) and then stress with Eq.~(\ref{eq:sevenc}).

\begin{figure}
\includegraphics[width=250pt]{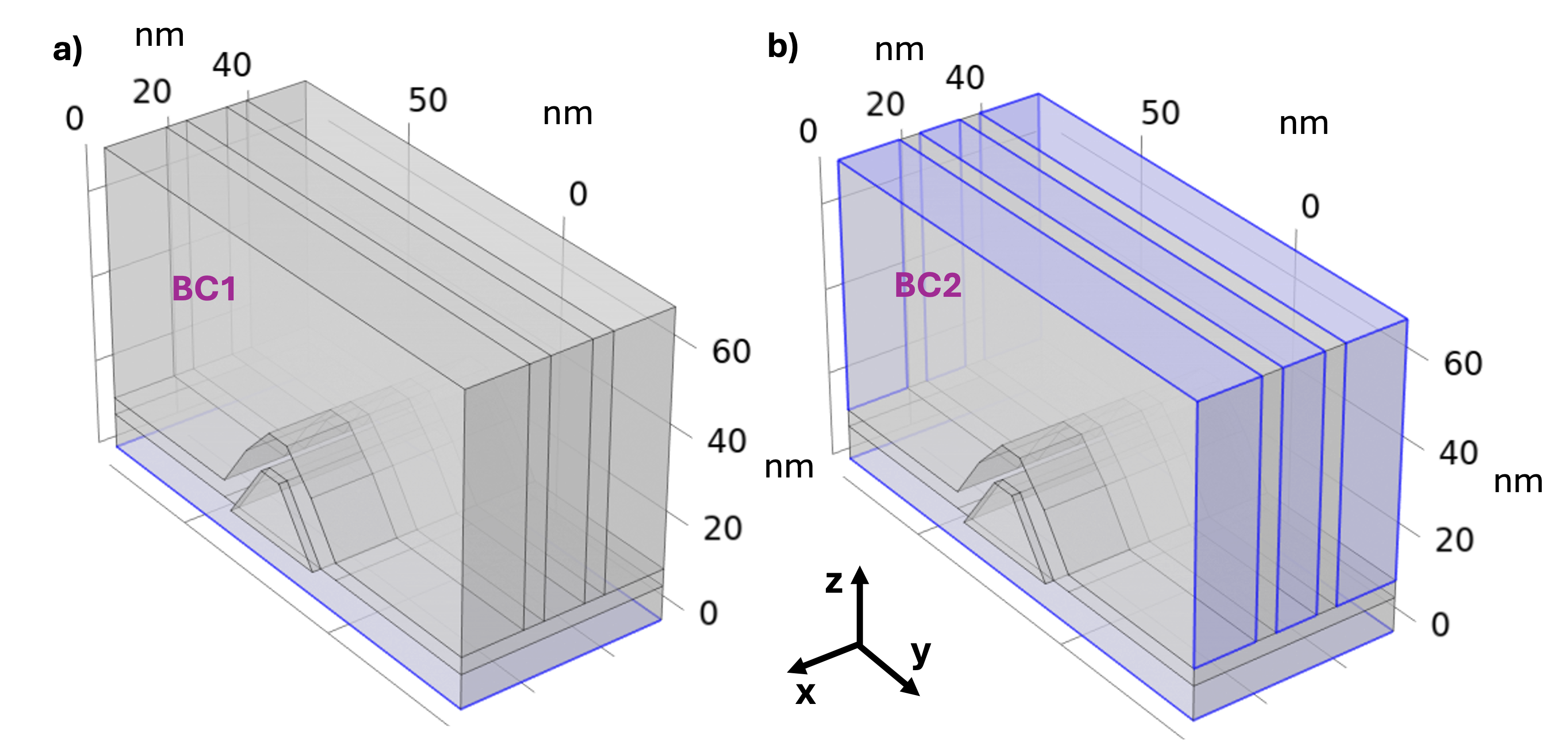}
\caption{\label{fig:BCs}Wire-frame representation of the quantum device structure from Fig.~\ref{fig:device} as constructed in COMSOL \cite{multiphysics1998introduction}. The blue surfaces denote the planes where rigid BCs are imposed. a) Fixed substrate (BC1). b) Fixed substrate and gate outer faces (BC2).}
\end{figure}

Here, we use the COMSOL multiphysics software \cite{multiphysics1998introduction} to perform finite element method (FEM) calculations of the strain tensor resulting from the cooling of the quantum device in Fig.~\ref{fig:device}. Since direct measurement of the local strain at cryogenic temperature is extremely challenging, if at all doable, it is not possible to determine which regions remain at fixed positions, i.e., are not affected by the cooling, and which ones undergo the largest contraction or expansion. We therefore devised two potential scenarios for the cooling-induced material displacements. Figure~\ref{fig:BCs} displays the corresponding device structures, as defined in COMSOL, where the transparent blue surfaces depict the regions where rigid BCs are imposed ($\boldsymbol{u}=0$). As compared to the geometry in Fig.~\ref{fig:device}, in the thermo-mechanical simulations, the TiN and SiO$_{2}$ parts of the FinFET are extended by 40 nm on each side to account for the thickness-dependent contraction of the gates on the SiO$_{2}$/Si stack below.

In both chosen configurations the dimensions of the buried substrate are assumed to be sufficiently large as compared to those of the fin to keep the bottom of the device structure fixed and use it as rigid BC. The behavior of the material stack (Si/SiO$_{2}$/TiN) mounted on the substrate then plays an important role. However, there is a priori no straightforward way to precisely extract the strain distribution inside the silicon channel after the quantum device being cooled down. To keep our analysis as general as possible, we conceive two “fictitious”, yet realistic cooling scenarios leading to either compressive or tensile strain inside the Si channel. In the first case, labeled BC1, we assume that the material stack can freely contract towards the substrate, which is fixed. In the second case (BC2), the external boundaries of the material stack corresponding to the surface of the TiN electrodes are fixed as well, thus preventing inward contractions of the SiO$_2$ and Si regions. In other words, the TiN gates act as anchor points and cause the expansion of the device interior. Combinations of experimental investigations and simulations of the strain-dependent qubit properties are needed to clearly identify the more realistic scenario.

Next, a strain tensor $\boldsymbol{\varepsilon}$ is calculated from the displacement field $\boldsymbol{u}$. It enters the $k\cdot p$ Hamiltonian through the Pikus-Bir terms, which take as input parameters the strain tensor components from Eq.~(\ref{eq:seven}) ($\varepsilon_{xx}$, $\varepsilon_{yy}$, $\varepsilon_{zz}$, $\varepsilon_{xy}$, $\varepsilon_{xz}$, and $\varepsilon_{yz}$) and the material parameters $a_{\nu}$, $b$, and $d$ known as the Pikus-Bir deformation potentials \cite{alma999115924401631}\cite{pollak1990effects}. 

\begin{subequations}
\label{eq:eight}
\begin{eqnarray}
P_{\varepsilon}=a_{\nu}(\varepsilon_{xx}+\varepsilon_{yy}+\varepsilon_{zz}), 
\end{eqnarray}
\begin{eqnarray}
Q_{\varepsilon}=-\frac{b}{2}(\varepsilon_{xx}+\varepsilon_{yy}-2\varepsilon_{zz}),
\end{eqnarray}
\begin{eqnarray}
R_{\varepsilon}=\frac{\sqrt{3}}{2}b(\varepsilon_{xx}-\varepsilon_{yy})-id\varepsilon_{xy},
\end{eqnarray}
\begin{eqnarray}
S_{\varepsilon}=-d(\varepsilon_{xz}-i\varepsilon_{yz}).
\end{eqnarray}
\end{subequations}

The $P_{\varepsilon}$, $Q_{\varepsilon}$, $R_{\varepsilon}$, and $S_{\varepsilon}$ values are added to the respective entries of the 6$\times$6 $k\cdot p$ Hamiltonian. As our SP solver relies on a FD grid, the FEM strain tensor from COMSOL must be first interpolated. Hence, after the static mechanical simulation, the strain tensor components are extracted at each point matching the FD grid and then used to compute the local entries of $P_{\varepsilon}$, $Q_{\varepsilon}$, $R_{\varepsilon}$, and $S_{\varepsilon}$. The simulation routine explained in Section \ref{subsec:General simulation} can be restarted with the previously converged electrostatic potential without strain as initial guess to speed up the numerical calculations. The SP loop is now evaluated in the presence of strain, which affects the qubit energy levels and orbital basis wave functions. 

\section{\label{sec:Results}Results \& Discussion}
\subsection{\label{subsec:Linear thermal contraction}Linear thermal contraction simulation}

\begin{table}[h]
\caption{\label{tab:table1}%
List of materials parameters used in the thermal and SP simulations: Young's modulus ($E$), thermal expansion coefficient ($\alpha$), Poisson's coefficient ($\nu$), Lüttinger's parameters ($\gamma_{1}$, $\gamma_{2}$, and $\gamma_{3}$), split-off spin-orbit energy ($\Delta_{0}$), isotropic magnetic parameter ($\kappa$), and deformation potentials ($a_{\nu}$, $b$, and $d$) [19][40]. }
\begin{ruledtabular}
\begin{tabular}{cccccccc}
&$E$&$\alpha\times10^{-6}$&$\nu$&
$\gamma_{1}$&$\Delta_{0}$&$\kappa$&$a_{\nu}$ \\
&[GPa]&[K$^{-1}$]&&$\gamma_{2}$&[eV]&&$b$ \\ 
&&&&$\gamma_{3}$ &&&$d$ \\ 
\hline
Si&169&2.6&0.27&4.285&0.044&-0.42&2.46\\
&&&&0.339&&&-2.35\\
&&&&1.21&&&-5.32\\
%\hline
SiO$_{2}$&73&0.49&0.17&-&-&-&-\\
%\hline
TiN&43&9.35&0.33&-&-&-&-
\end{tabular}
\end{ruledtabular}
\end{table}

The parameters used to perform the thermo-mechanical simulations of the FinFET's material contractions are listed in the first four columns of Table~\ref{tab:table1} for Si (channel and substrate), SiO$_{2}$ (oxide), and TiN (metallic electrodes). In our analysis, the materials are described with isotropic and temperature-invariant coefficients so that the Young modulus and the thermal expansion coefficient take single values. Moreover, in our model we neglect residual strain that might be induced during the fabrication process. Consequently, the thermal contractions/expansions are calculated under the assumption of no pre-existing strain, i.e, the TiN/SiO$_2$ and SiO$_2$/Si interfaces are assumed to be perfectly relaxed at room temperature (300K) before cooling them down to 4K.

\begin{figure}[h]
\includegraphics[width=250pt]{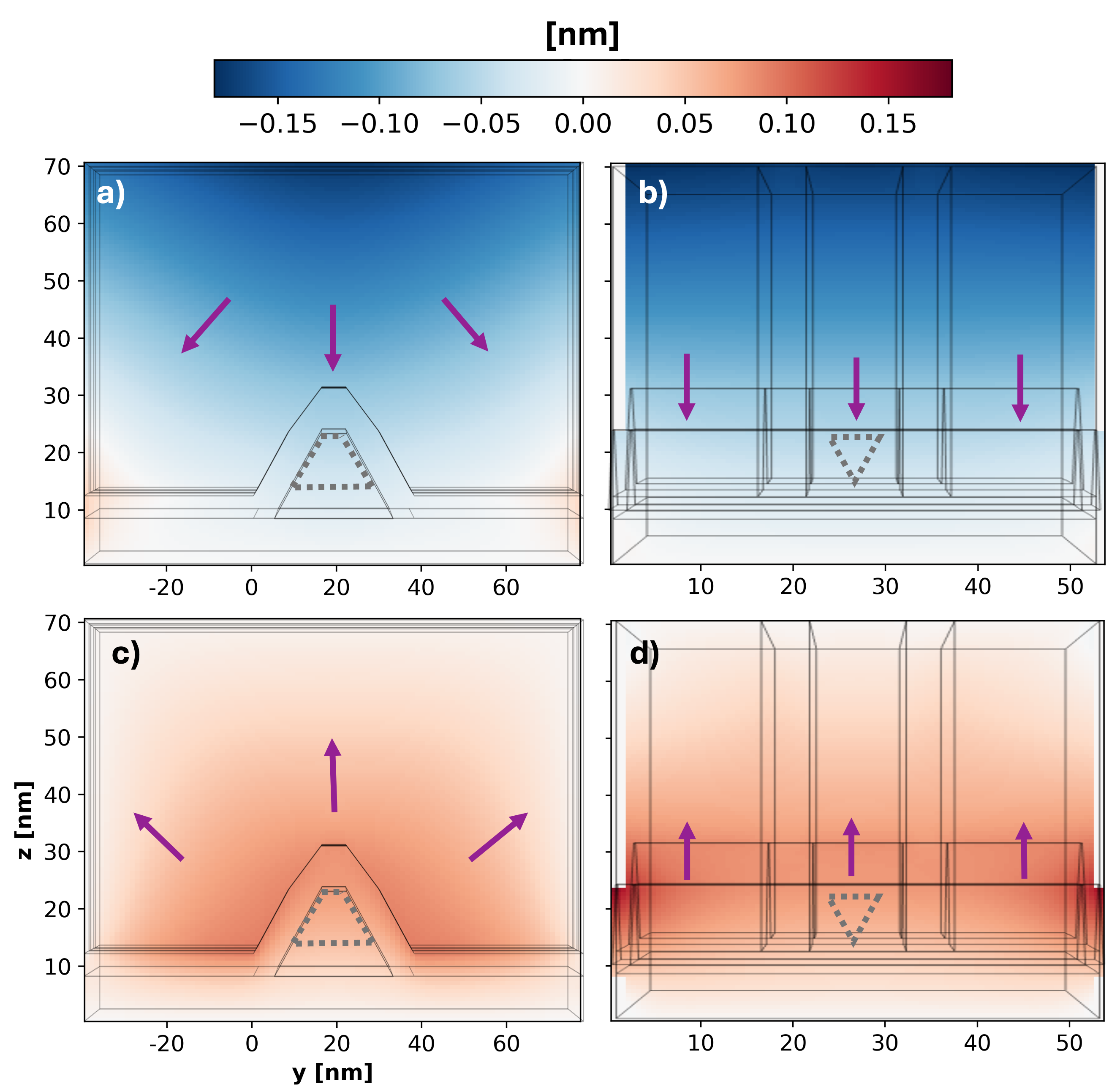}
\caption{\label{fig:disp} Colormaps of the displacement field $u_z$ in the plane formed by the device cross section ($yz$- plane) from Fig.~\ref{fig:device} at $x$ = 27 nm for BC1 a) and BC2 c), and along the $xz$-plane corresponding to the transport direction at $y$ = 19 nm for BC1 b) and BC2 d). The solid black lines denote the device’s edges. The gray dashed trapezoids represent the approximate position of the qubit and the purple arrows the direction of the thermally-induced forces.}
\end{figure}

The impact of the thermal-induced deformation on the displacement field $\boldsymbol{u}$ from Eq.~(\ref{eq:sevend}) is illustrated in Fig. \ref{fig:disp} where $u_z$ ($z$-component of the displacement field) is plotted in the plane corresponding to the device cross-section ($yz$-plane) at $x=27$ nm in a) and c) for the BC1 and BC2 cases, respectively. It is then reported along the transport direction ($xz$-plane) at $y=19$ nm in b) and d). The $u_x$ component (displacements along $x$) in the plane perpendicular to the channel ($yz$-plane) and the $u_y$ component (displacements along $y$) in the plane parallel to the channel ($xz$-plane) have negligible values owing to the symmetry of the system (i.e, geometry and BCs). Moreover, in the BC1 case the deformation originating from the contraction of the gates and contacts has a lower impact at the dot's position for the $u_y$ and $u_z$ components of the displacement field $\boldsymbol{u}$. However the $u_y$ component in the BC2 case has a noticeable effect at the dot's position. This will be later discussed in the context of the strain tensor. In addition, the exact same thermo-mechanical strain simulations were performed on a slightly larger cross-section that better matches the dimensions of the original quantum device from \cite{camenzind_hole_2022}. The resulting displacement maps (not shown) exhibit very similar strain features, both qualitatively and quantitatively. Hence, the applied reduction of the structure dimensions allows to decrease the computational burden, without significantly affecting neither the thermo-mechanical nor the electronic ones.

\begin{figure}[h]
\includegraphics[width=220pt]{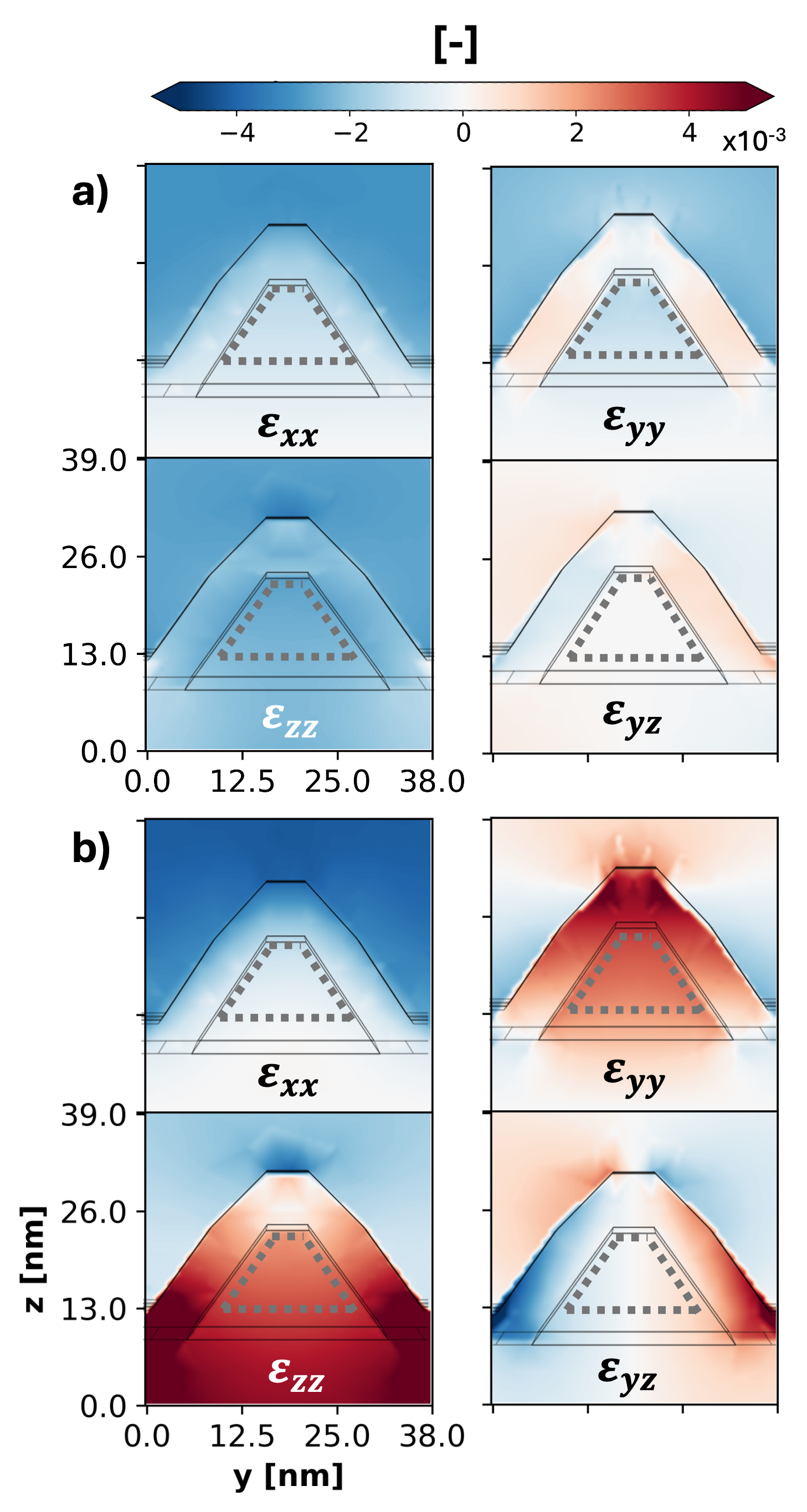}
\caption{\label{fig:strain} Colormaps of the strain tensor components $\varepsilon_{xx}$, $\varepsilon_{yy}$, $\varepsilon_{zz}$, and $\varepsilon_{yz}$ in the $yz$-plane at $x$ = 27 nm in Fig. ~\ref{fig:device} for a) BC1 and b) BC2. The solid black lines denote the cross-section edges and the gray dashed trapezoids enclose the approximate position of the qubit.}
\end{figure}

From the displacement maps in Fig.~\ref{fig:disp}, the influence of the rigid BCs can be clearly identified (white regions) at the bottom of the structure for BC1 in sub-plots a) and c), or all around it for BC2 in sub-plots b) and d). In these regions, the displacements reach their minimum value. We first observe that a change of BCs between BC1 and BC2 switches the direction of the $u_z$ displacement inside the device. Under BC1 the downward pointing arrows indicate that the whole material stack tends to be compressed toward the fixed substrate while the upward pointing arrows in BC2 show that the material stack expands away from the substrate. Furthermore, due to its comparatively higher thermal expansion coefficient and the larger thickness of the TiN layers, the metallic gates undergo the largest deformation. The displacement field observed in TiN is indeed strongly inhomogeneous and ranges from -0.18 nm to 0.03 nm for BC1 and from 0 to 0.16 nm for BC2. The deformation gradient propagates to the lower-lying materials with smaller thermal expansion coefficient (i.e., Si and SiO$_{2}$), which leads to a varying displacement field component $u_z$ from the top of the SiO$_{2}$ layer to the bottom of the Si channel ranging from -0.03 nm to 0.015 nm for BC1 and from 0.07 nm to 0.025 nm for BC2. The BC layout determines the localization the maximum and minimum deformation in the quantum device. BC1 with fixed substrate induces a contraction of the device with a maximum displacement of -0.18 nm at the uppermost boundary and approximately -0.045 nm at the dot position. In BC2 where additionally the outer TiN gate planes are fixed the interior of the device expands and the maximum displacement is shifted to the outer corners at the SiO${_2}$/TiN interface (0.09 nm) and to the upper boundaries of the source and drain (0.16 nm). At the dot position, an overall displacement of approximately 0.065 nm is observed.

Using Eq.~(\ref{eq:seven}) the components of the strain tensor $\boldsymbol{\varepsilon}$ inside the Si fin can be calculated from the gradient of the displacement field $\boldsymbol{u}$. The results are shown in Fig.~\ref{fig:strain} for the device cross-section in the middle of the channel, at $x$ = 27 nm. For clarity the approximate location of the qubit is enclosed by gray dotted trapezoids. The downward motion of the material stack observed under BC1 translates into a compressive thermal stress (negatively valued blue regions) dominated by $\varepsilon_{zz}$ ($\approx-2.5\times$10$^{-3}$), with a smaller contribution of $\varepsilon_{xx}$ and $\varepsilon_{yy}$ ($\approx-1\times10^{-3}$). We note the uniform contraction of the metallic gates along the diagonal axes of the strain tensor ($\varepsilon_{xx}$, $\varepsilon_{yy}$, and $\varepsilon_{zz}$) in the BC1 scenario. At the same time, the positive $\varepsilon_{yy}$ strain component observed in the SiO$_2$ layer indicates that the change in thermal expansion coefficient at the Si/SiO$_2$ interface (from $\alpha_{Si}=2.6\times10^-6$ $K^-1$ to $\alpha_{SiO_2}=0.49\times10^-6$ $K^{-1}$) induces tensile strain in the oxide layer along the $y$ axis. Under BC2, on the other hand, the Si fin experiences a tensile stress (positively valued red regions), with a significant contribution from both $\varepsilon_{zz}$ ($\approx3\times$10$^{-3}$) and $\varepsilon_{yy}$ ($\approx2.5\times$10$^{-3}$). This traction originates from the outward motion of the surrounding metallic gates (blue regions) at the TiN/SiO2 interface. Simultaneously, the metallic gate experiences an overall displacement upwards along the $z$ axis (blue $\varepsilon_{zz}$) as well as opposite lateral displacements along the $y$ axis over the Si fin height (blue $\varepsilon_{yy}$) . This leads to a traction of the SiO2 layer in the $yz$-plane (tensile strain) that further propagates to the Si channel. Moreover, a small compressive stress ($\approx-0.7\times10^{-3}$) coming from the device source and drain contraction is found for $\varepsilon_{xx}$ as well as a non-vanishing $\varepsilon_{yz}$ term having mixed compressive and tensile stress ($\pm\approx1\times10^{-3}$). Because of the aforementioned symmetry arguments, the other mixed components of the strain tensor are negligible along the high symmetry planes for both BC1 and BC2.

\subsection{\label{subsec:Quantum dot formation}Quantum dot formation}
\begin{figure*}[t]
\includegraphics[width=400pt]{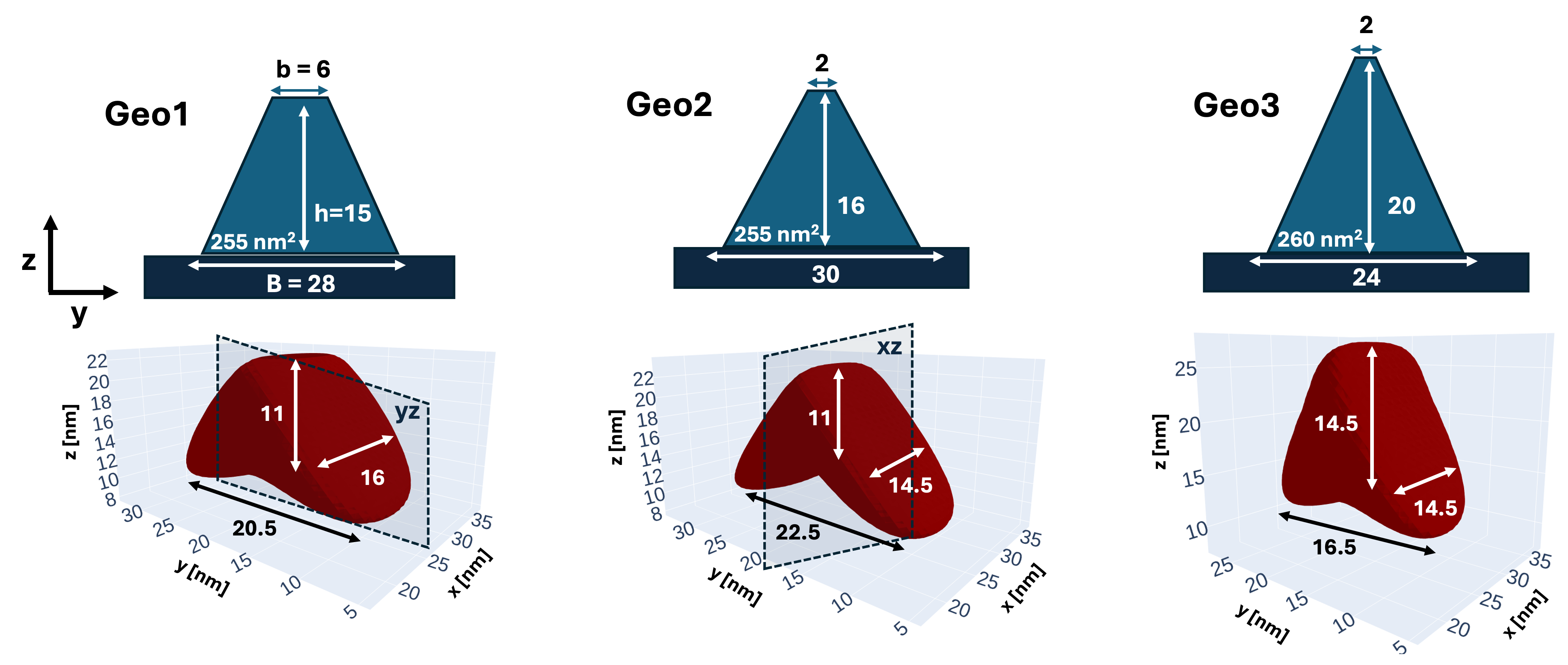}
\caption{\label{fig:Geo_Iso}Top: Shape and dimensions of the three different, strain-free FinFET's cross sections investigated here and labeled Geo1, 2, and 3. Their height $h$ and bottom (top) width $B$ ($b$) vary, with increasing $h/B$ and decreasing $b/B$ ratios from Geo1 to Goe3. Bottom: Corresponding ground state isodensity surfaces within which 90\% of the total qubit charge is encapsulated. The dimensions of these isosurfaces along the $x$, $y$, and $z$ axes are indicated as double arrows. All three isodensites possess two mirror planes, $yz$ and $xz$, denoted by the blue dashed rectangles in the left and central 3D plots.}
\end{figure*}

\begin{figure*}[t]
\includegraphics[width=450pt]{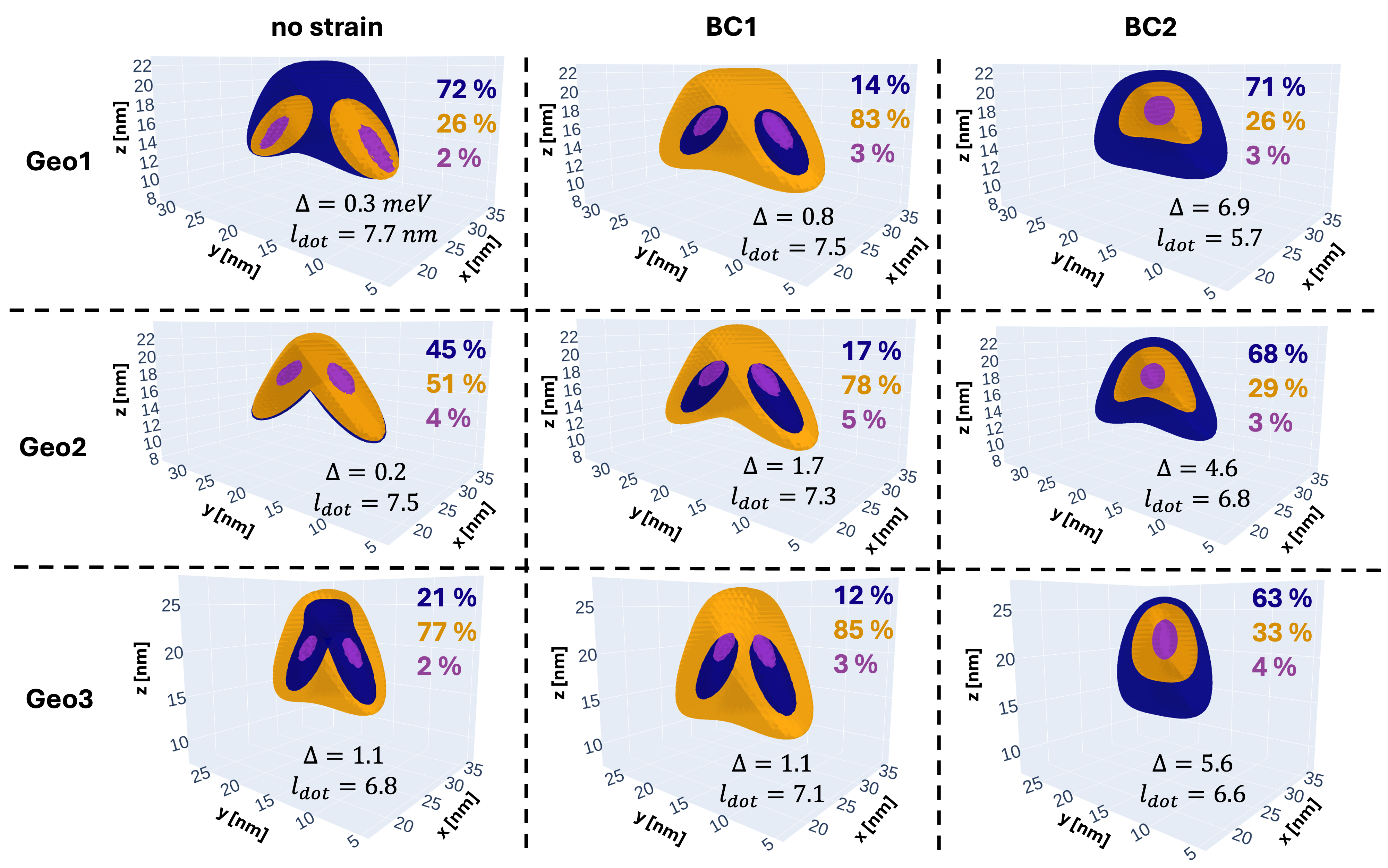}
\caption{\label{fig:band_mixing} Isosurfaces of the heavy-hole (blue), light-hole (orange), and spin-orbit (purple) contributions to the square of the ground state charge density in the unstrained (left), BC1 (middle), and BC2 (right) cases. The resulting band mixing energy spacing ($\Delta$ [meV]) and equivalent dot's length ($l_{dot}$ [nm]) are indicated in each sub-plot.}
\end{figure*}

To compute the qubit ground state, SP simulations are performed with the Si parameters listed in Table~\ref{tab:table1}, i.e, the Lüttinger’s parameters, the spin-orbit split-off energy, and the Pikus-Bir’s deformation potentials. The discretized strain tensor from Fig.~\ref{fig:strain} along with the deformation parameters are supplied to the SP solver to incorporate the local influence of strain. The formation of the ground-state is studied for three different device geometries presented in Fig.~\ref{fig:Geo_Iso} and labeled Geo1 to Geo3. In these structures, the longitudinal dimensions are kept unchanged ($L_{c}=50$, $L_{lead} = 15$, $L_{gap} = 5$, in nm) and the cross-section areas are set to approximately the same value ($\approx 260$ nm$^{2}$). The ratio between the two bases $b$ (top) and $B$ (bottom) and the height $h$ varies so that $h/B$ ($b/B$) increases (decreases) from Geo1 to Geo3. Systematic comparisons are conducted between the electrostatically induced QD under strain-free conditions and in the presence of thermally strained deformations with the BCs of Fig.~\ref{fig:BCs}, all at 4K. 

In the Geo1 device, a hole qubit is formed by applying a bias V$_{L,R}$ = 0.78 V to the side gates and a bias V$_{P}$ = 0.38 V to the plunger gate. For Geo2 and Geo3, the following bias pairs were used to induce a QD: V$_{L,R}$ = 0.84 V / V$_{P}$ = 0.33 V and V$_{L,R}$ = 0.84 V / V$_{P}$ = 0.36 V, respectively. The bias conditions to create a QD remain the same with and without strain with only a shift in the qubit ground state energy. The qubit strain-free shapes are compared in Fig.~\ref{fig:Geo_Iso} by plotting the ground state isodensity surfaces that enclose 90\% of the QD's probability density function (square of the wave function). Based on the spatial profile of the isodensities, it can be seen that the device cross section has a strong influence on the QD shape. For example, the close to one $h/B$ ratio of Geo3 leads to an elongated density along the $z$ axis (14.5 nm) with a smaller spread along the $y$ direction of 16.5 nm at the triangle base. On the other hand, the intermediate $h/B$ ratio of Geo2 gives rise to an isodensity that predominantly spreads parallel to the Si/SiO$_{2}$ interfaces of the structure, reaching 22.5 nm lateral amplitude along the $y$ axis. This produces a depleted region in the lower part of the channel and a dot with a narrower extent along the $z$ direction (11 nm). In Geo1 the smaller $h/B$ and larger $b/B$ ratios allow for the QD to further extend along the $x$ direction (16 nm vs. 14.5 nm for Geo2 and Geo3). This reduces the volume of the depleted region in the lower part of the structure and generates a QD showing less spatial anisotropy than the two other geometries.

The impact of strain on the qubit ground-state can be better appreciated by examining the different components of its squared wave function, i.e., the band mixing plotted in Fig.~\ref{fig:band_mixing}. In each sub-plot, the isosurfaces of the HH are drawn in dark blue, the LH ones in orange, and the SO contributions in purple. Besides, for each geometry/strain configuration we extract the energy splitting $\Delta$ (in meV) between the ground-state and first excited state and the equivalent dot length $l_{dot}$ (in nm). Firstly, when a relaxed structure is considered at 4K (no strain), the band mixing displays a strong structure-dependent behavior since the geometry dictates the electrostatic potential landscape in which the qubit operates. When the $h/B$ ratio is close to one (Geo3) the LH dominates (HH=21\%/LH=77\%/SO=2\%), while a small $h/B$ and larger $b/B$ ratio (Geo1) favors the HH population (HH=72\%/LH=26\%/SO=2\%). A similar behavior was observed in the theoretical study of Si and Ge nanowires where the presence of two strong directions of confinement enhances the band mixing and gives rise to a dominating LH population \cite{csontos_between_nodate,pedersen_energy_1996,moratis_light-hole_2021,kloeffel_direct_2018}. On the other hand, in 2D-like systems where there is only one single direction of confinement (usually the $z$ direction) a predominantly HH-like ground-state is expected \cite{martinez_hole_2022,venitucci_electrical_2018,venitucci_simple_2019}. Furthermore, the weaker confinement of Geo1 and Geo2 translates into a smaller splitting energy (0.3 meV and 0.2 meV, respectively vs. 1.1 meV for Geo3) and a greater dot length (7.7 nm and 7.5 nm, respectively vs. 6.8 nm for Geo3). 

While only small modifications of the qubit shape are observed when strain is applied, the major change comes from the renormalization of the HH, LH, and SO contributions to the QDs ground-state. Regardless of the geometry, for BC1 (compressive strain) the LH population dominates whereas for BC2 (tensile strain) the HH population is more important. A rapid analysis confirms the validity of these results if, in first approximation, we only consider the predominant $\varepsilon_{zz}$ component of the strain tensor and the fact that the effective masses for the LH ad HH are close (for $\langle100\rangle$, $m^{*}_{LH}=0.204$ and $m^{*}_{HH}=0.274$ \cite{donetti_hole_2011}). We find from the Pikus-Bir formulations in Eqs.~(\ref{eq:eight}a) and (\ref{eq:eight}b) that a negative $\varepsilon_{zz}$ (BC1 case) leads to a downward (upward) energy shift of the $P+Q$ ($P-Q$) term and hence, shifts the HH (LH) contribution down (up) in energy so that the ground-state has a stronger LH character. Similarly, a positive $\varepsilon_{zz}$ (BC2 case) switches the direction of the shift and creates the opposite trend. Additionally, we observe that the computed ground-states under BC2 retain a greater proportion of the minority LH population than BC1 does with the minority HH population. For instance in Geo2 LH=29\% under BC2 and HH=17\% under BC1. This phenomenon can be traced back to the non-vanishing $\varepsilon_{yz}$ component of the strain tensor in Fig.~\ref{fig:strain}b) (lower right map), which contributes to the $S$ term of the Pikus-Bir Hamiltonian in Eq.~(\ref{eq:eight}d) . This term admixes the LH and HH populations, thus explaining the greater LH population observed in BC2.

Lastly, the extraction of the ground and first excited states reveals that their energy separation $\Delta$ is far greater under tensile strain (BC2) than it is in case of compressive strain (BC1). For example, for Geo1, $\Delta$ increases from 0.8 meV to 5.6 meV when going from BC1 to BC2, while it was only 0.3 meV in the relaxed structure. This in turn leads to a stronger confinement and more localized QDs in the BC2 scenario, as can be noticed from the $l_{dot}$ values. Hence, if the TiN gate electrodes could only marginally contract during the cooling of the FinFET device to cryogenic temperatures, the resulting quantum dots would be more confined and the ground-state energy better separated from the first excited one.

\subsection{\label{subsec:qubit Larmor and Rabi}qubit Larmor and Rabi frequencies}

\begin{figure}
\includegraphics[width=250pt]{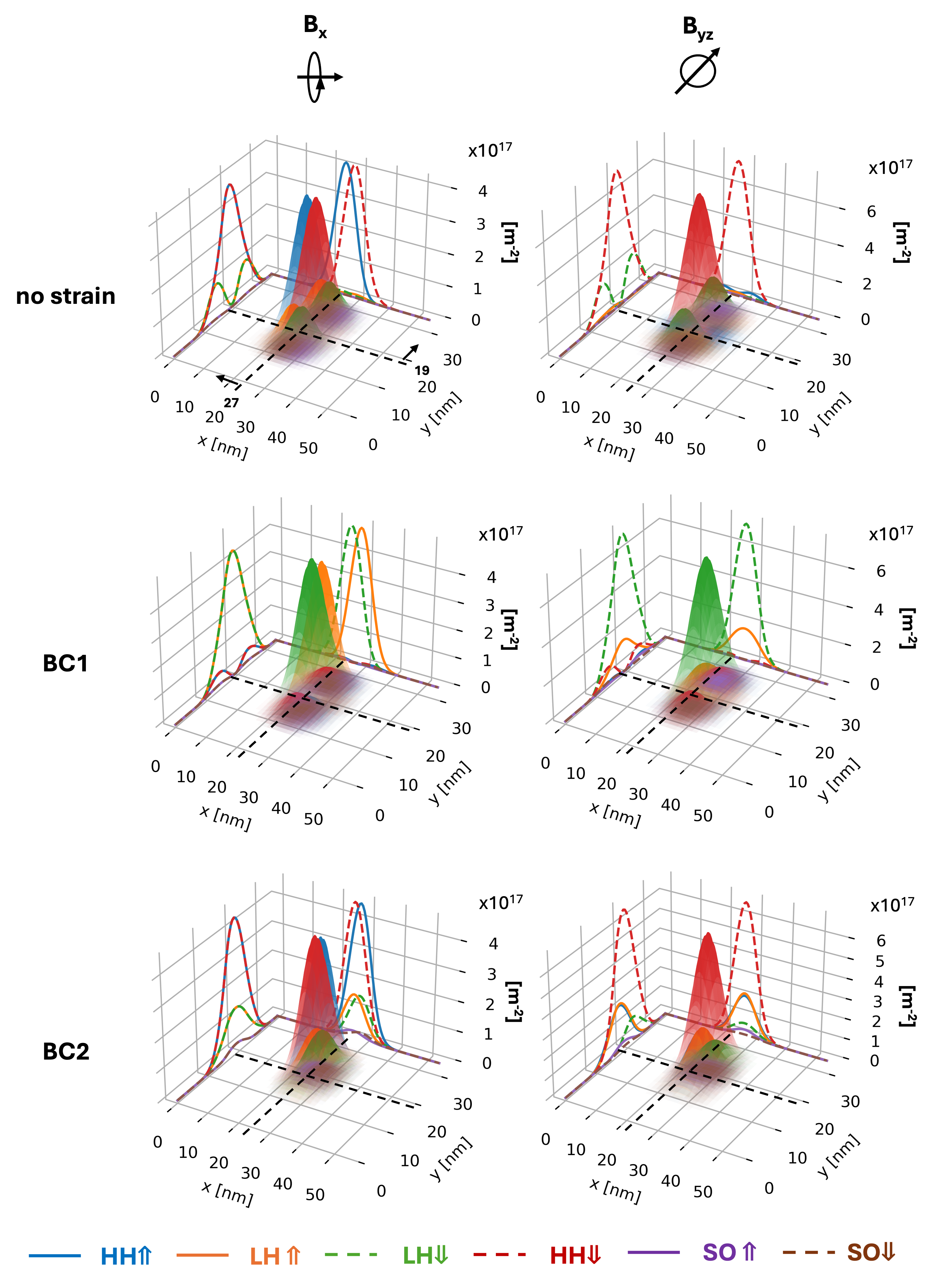}
\caption{\label{fig:spin_mixing}Representation of the qubit’s lowest energy state in Geo1 under a static magnetic field ($B$ = 100 mT) oriented along the $x$ and $yz$ direction for the unstrained, BC1, and BC2 cases. The central surface plots represent the qubit’s charge density in cm$^{-2}$ averaged along the $z$ axis. The 1D curves drawn at $x$ = 0 and $y$ = 42 nm are the charge densities along the dashed lines passing through the dot’s center, i.e, at $x$ = 27 nm and $y$ = 19 nm. The following color scheme is used for the 6 basis states: blue (HH$\Uparrow$), orange (LH$\Uparrow$), green (LH$\Downarrow$), red (HH$\Downarrow$), purple (SO$\Uparrow$), and brown (SO$\Downarrow$).}
\end{figure}

Next, the influence of homogeneous magnetic fields on the qubit Larmor and Rabi frequencies is studied. This is done by adding $\kappa$-dependent terms to the diagonal of the discretized Hamiltonian matrix corresponding to the device of interest as well as a vector potential to its non-diagonal blocks (see Appendix \ref{AppendixA} for more details). SP simulations are then performed with this modified Hamiltonian. The previously converged electrostatic potential obtained at $B$ = 0 T is used as initial guess for the runs with an applied magnetic field $B$ = 100 mT. This magnetic field couples the spin-orbit basis states of the $k\cdot p$ Hamiltonian and lifts the degeneracy of the qubit ground-state by opening an energy gap (Zeeman splitting). The latter is linearly proportional to the applied $\boldsymbol{B}$ field and to the effective $g$-factor $g^{*}$ (see Eq.~\ref{eq:three}). In Fig.~\ref{fig:spin_mixing}, the ground-state wave function of Geo1 is decomposed into its six components (HH$\Uparrow$, LH$\Uparrow$, SO$\Uparrow$, HH$\Downarrow$, HH$\Downarrow$, and SO$\Downarrow$), under different strain configurations (no strain, BC1, and BC2) and magnetic field orientations ($B$ = [0.1 0 0] T and $B$ = [0 0.07 0.07] T). The corresponding two-dimensional charge densities (averaged over the $z$ direction) are represented as well. Additional plots along the remaining magnetic directions are provided in \ref{Appendix4}.

\begin{figure*}[t]
\includegraphics[width=450pt]{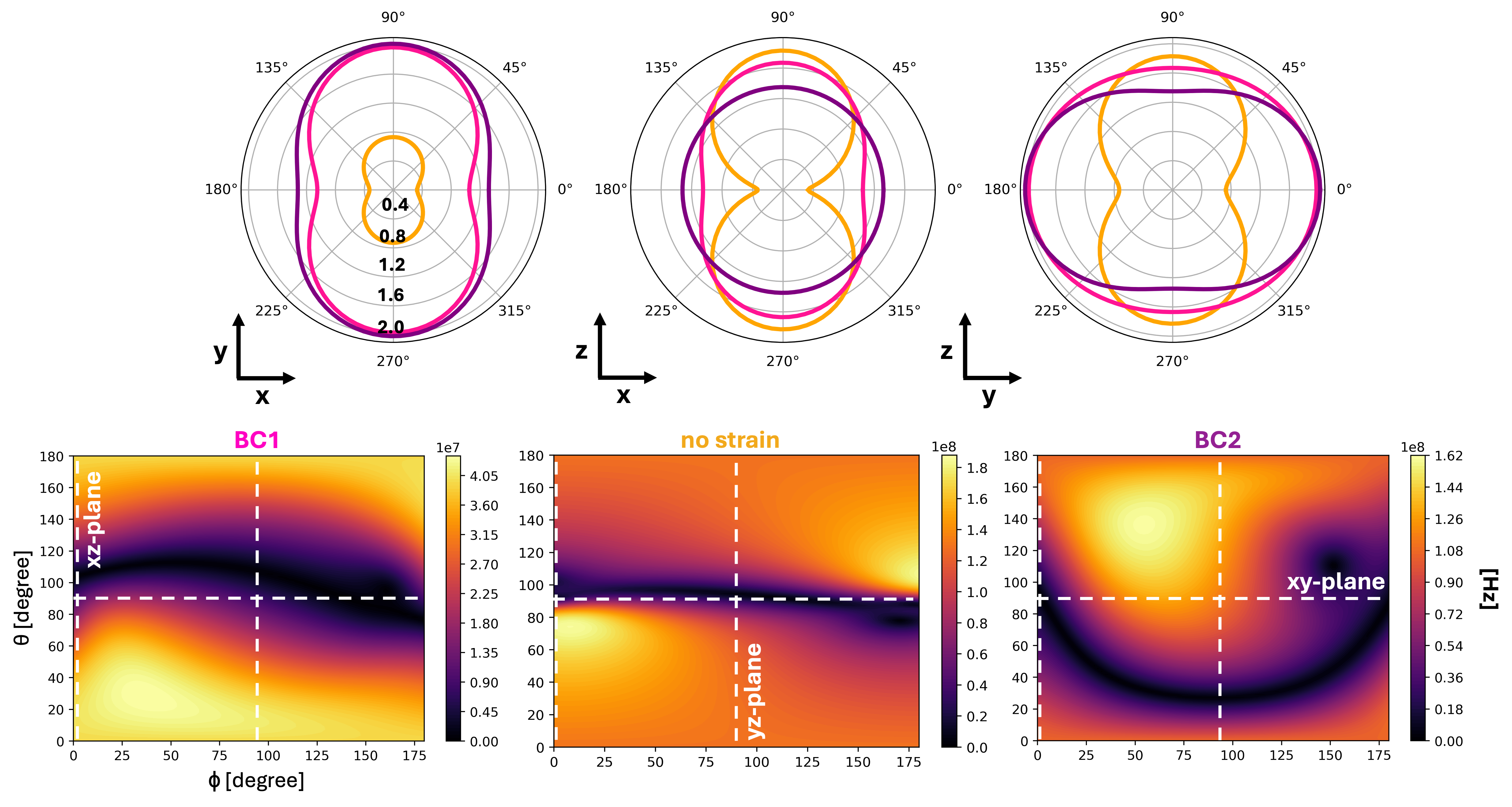}
\caption{\label{fig:g_Rabi} Top: Cross sections of the $g$-tensor along the $xy$-, $xz$-, and $yz$-plane for the unstrained (orange), BC1 (pink), and BC2 (purple) cases. Bottom: Rabi frequency maps as a function of the azimuthal ($\theta$) and polar ($\phi$) angles of the magnetic field. The Rabi frequencies are computed for Geo1, at a constant Larmor frequency $f_{L}$ = 8 GHz and with $V_{ac}$ = 1 mV applied to the plunger gate. The $xy$-, $xz$-, and $yz$-plane are marked by the corresponding white dashed lines}
\end{figure*}

Without strain, the HH population dominates (72\%), with equal contributions from both spin orientations in case of out-of-plane field $B_{x}$, but a stronger contribution from the $\Downarrow$ spin orientation when the magnetic field is in-plane ($B_{yz}$). The analysis of the total angular momentum matrices in Eq.~(\ref{eq:two}) provides a clearer picture of the spin mixing inside the qubit. Along the equivalent [100] directions, $J_x$ and $J_y$ share the same matrix entries, namely the one coupling both the $\Uparrow$ and $\Downarrow$ populations. On the other hand, $J_z$ is diagonal and it splits the $\Uparrow$ and $\Downarrow$ populations. Since $g_{z}>g_{y}$ (see Fig.~\ref{fig:g_Rabi}) $B_{yz}$ greatly reduces one of the orientations. Equivalently, the same argument can be employed to explain the strain cases. As already stated before, the nature of strain determines the majority carrier (i.e., LH for BC1 and HH for BC2), whereas the orientation of the magnetic field governs the distribution of the spin-orbit basis states in the qubit ground state (i.e., $\Uparrow$ and $\Downarrow$ for $B_{x}$ and predominantly $\Downarrow$ for $B_{yz}>$0).

The extraction of the Zeeman energy is straightforward. It consists of measuring the energy spacing between the previously doubly-degenerate ground states. Hence, by performing SP simulations using six independent magnetic directions it is possible to reconstruct the entire qubit $g$-tensor \cite{crippa_electrical_2018}. Nevertheless, taking advantage of the linear dependence of the system on $\boldsymbol{B}$, we only need to perform a first-order linear expansion of the magnetic operator along the $x$, $y$ and $z$ axes. This computationally less intensive approach was proposed in \cite{venitucci_electrical_2018} to create the $g$-tensor from a single calculation of the qubit ground-state. The polar plots of Geo1's $g$-tensor along the $xy$-, $xz$- and $yz$-plane are presented in the top panels of Fig.~\ref{fig:g_Rabi} for the different strain cases. Regardless of the configuration, the principal magnetic axes remain oriented along the devices principal axes (i.e, $X=[100]$ , $Y=[010]$, and $Z=[001]$) which reflects the symmetry of the applied strain and of the device itself. Under relaxed conditions (orange curves) the $g$-tensor takes the principal values $g_{x}$ = 0.33, $g_{y}$ = 0.73, and $g_{z}$ = 1.83, exhibiting a strong three-dimensional anisotropy with a peanut shape in all three planes ($xy$, $xz$, and $yz$). This result underlies the dependence of the $g$-tensor on the device cross section and thus on the confinement potential since in Geo1 the confinement increase from $x$ to $z$. However, the application of strain tends to reduce the $g$-tensor anisotropy through an uneven increase of the principal $g_{x}$ and $g_{y}$ factors. From their unstrained values of 0.33 and 0.73, $g_{x}$ and $g_{y}$ increase to 1.05 and 1.97 for BC1 (pink curves) and to 1.32 and 1.35 for BC2 (purple curves), respectively.

The corresponding Rabi frequencies (RF) are calculated according to Eq.~(\ref{eq:five}) where $g'(V_{0})$ is obtained from the self-consistent simulations near the reference plunger gate potential $V_{0}$ (i.e., at $V = V_{0} \pm\delta V$). The resulting RF are given in the three bottom panels of Fig.~\ref{fig:g_Rabi} for the relaxed (orange), BC1 (pink) and BC2 (purple) cases. The gate potential $V_{p}$ is modulated by a microwave signal of amplitude $V_{ac}$ = 1 mV at constant Larmor frequency $f_{L}$ = 8 GHz. The $\theta$ ($\phi$) angle represents the polar (azimuthal) direction of the magnetic field vector. The maps of the RF show a non-monotonic behavior as a function of the magnetic field orientation. This results from the prismatic-like shape of the investigated qubit. In fact, the triangular cross-section confers an inherent anisotropy to the qubit envelope function, which translates into $\theta$- and $\phi$-dependent RF magnitudes (e.g., maximum of 4 MHz observed for $\phi$=30$^\circ$ and $\theta$ = 40$^\circ$ for the BC1 case). Localised RF maxima could also be observed in FinFET-like geometries in \cite{crippa_electrical_2018}, while RF maps for homogeneous/planar dots display only $\theta$-dependence (constant RF magnitude at fixed $\theta$ and varying $\phi$) \cite{abadillo-uriel_hole-spin_2023,venitucci_electrical_2018}.

Following the same symmetry arguments as in \cite{venitucci_electrical_2018}, we can use the qubit mirror planes to understand and explain the observed trends. In the absence of strain, the qubit possesses quasi-symmetric planes along the $yz$- and $xz$-plane drawn in Fig.~\ref{fig:Geo_Iso} that bring the RF to zero at $\phi$=0 or 90$^\circ$ and $\theta$ = 90$^\circ$ and extend into a region of vanishing RF in the vicinity of the $xy$-plane (i.e., extinction region). Due to the triangular shape of the fin the volume occupied by the dot slightly changes when the dot is being driven by the oscillating electromagnetic field, which leads to the observed modulation of the extinction region area along $\phi$.

The situation under thermally-induced strain is significantly different. We identify two features impacting the RF of strained qubits, namely the spatial distribution of the spin-orbit states and of the non-homogeneous strain around the qubit location. While strain only slightly alters the qubit shape, the spatial distribution of the spin-orbit basis states undergoes larger modifications (see Fig.~\ref{fig:spin_mixing}). In both cases, an applied magnetic field along the $x$ axis mixes both spin orientations, but it does not perturb the spin symmetry of the wave function, with a nearly zero RF along that direction as a consequence. For BC1 we see that compressive strain preserves the symmetry of the spin-orbit basis states for $B_{yz}$. Therefore, this more symmetric dot widens the extinction area of the RF and leads to its global decrease. In contrast, the HH$\Uparrow$ and LH$\Downarrow$ states in BC2 display an anisotropic spatial distribution when $B_{yz}$ is applied, inducing a global increase of the RF. 

Secondly, the thermal-elastic simulations reveal that the qubit experiences non-homogeneous strain along the different device axes which explains the bending of the extinction region observed in both the BC1 and BC2 cases. In fact, from the displacement (Fig.~\ref{fig:disp}) and strain maps (Fig.~\ref{fig:strain}) we notice that the strain components in the high-symmetry planes change with the height and width of the fin. Moreover, the calculation of the strain tensor in cross-sections away from the $yz$ mirror plane (see \ref{Appendix5}) shows that its main components as well as its now non-vanishing cross-terms ($\varepsilon_{xy}$ and $\varepsilon_{xy}$) vary in the directions along which the qubit extends. These variations are captured by $g’$ when the qubit envelope function is modulated by electromagnetic waves and its different regions are subject to changing strain gradients. 

\subsection{\label{subsec:Comparison with experimental device}Comparison with experimental device}

\begin{table}[h]
\caption{\label{tab:table2}%
Comparison of the device metrics obtained from self-consistent SP simulations and from measurement for dot Q1 and Q2 \cite{camenzind_hole_2022}. The dot length ($l_{dot}$) and the energy splitting ($\Delta$) are given in nm and meV, respectively. The $g$-factors ($g^*$) are given as $g_{x}/g_{y}/g_{z}$, and the Rabi frequency ($f_{R}$) in MHz for $\theta=\phi=90^\circ$ at constant $f_L=3$ GHz and $V_{ac}=1$ mV.}
\begin{ruledtabular}
\begin{tabular}{cccccc}
&$l_{dot}$&$\Delta$&
$g^*$&$f_{R}$ \\

\hline
No strain&6.9&0.8&2.05/0.22/0.56&7.5\\ 
%\hline
BC1&7.0&2.6&1.22/2.24/2.17&8.5\\
%\hline
BC2&7.0&2.7&1.30/2.06/2.23&104\\
%\hline
Exp. Q1&5.6&5.3&1.50/1.94/2.50&23\\
%\hline
\textbf{Exp. Q2}&7.1&3.3&1.60/2.35/2.70&10\\
\end{tabular}
\end{ruledtabular}
\end{table}

\begin{figure*}[t]
\includegraphics[width=450pt]{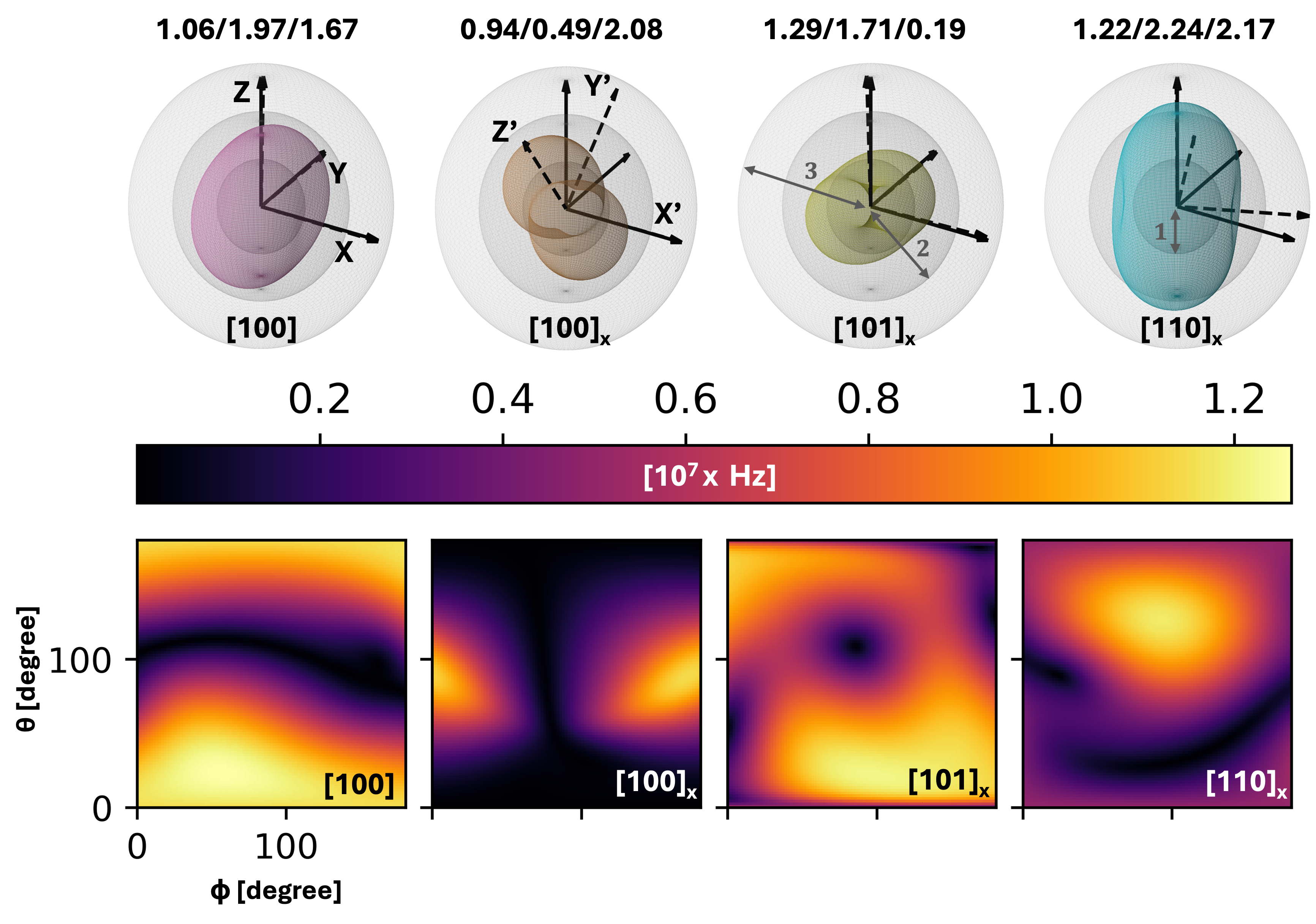}
\caption{\label{fig:comparison} Top: Principal $g$-factor values given as $g_{x'}/g_{y'}/g_{z'}$ along with their 3D $g$-tensor plots for the $[100]$ (pink), $[100]_x$ (brown), $[101]_x$ (green), and $[110]_x$ (cyan) directions under BC1. The dotted arrows represent the principal magnetic axes and the solid ones the device axes. Bottom: Corresponding Rabi frequency maps computed for Geo1, at a constant magnetic field $B$ = 1 mT and with $V_{ac}$ = 1 mV.}
\end{figure*}

Finally, in Table \ref{tab:table2} we compare the device metrics computed with our in-house SP solver \cite{luisier_full-band_2008} against those extracted from the experiments of Ref.~\cite{camenzind_hole_2022}. To reproduce the crystallographic orientation of the original experimental set-up (i.e, $x=[110]$, $y=[1\Bar{1}0]$, and $z=[001]$), the $6\times6$ $k\cdot p$ Hamiltonian was rotated as described in \cite{winkler_spin-orbit_2003}. Moreover, additional care was taken when computing the device strain tensor since the Si Young modulus is direction-dependent. Henceforth, the complete $6\times6$ compliance matrix $C$ (see Eq.~(\ref{eq:sevenc})) for the desired orientation is given to COMSOL as input of the thermal-elastic simulations \cite{zhang_anisotropic_2014}. 
For this new crystallographic orientation the bias conditions to create exactly one hole below the plunger gate were adjusted to the following voltages: $V_p = 0.40$ V and $V_{L,R} = 0.82$ V. Note that the rotation of the crystallographic axes involves a significant increase in the HH effective mass (for $[110]$: $m^{*}_{LH}=0.149$ and $m^{*}_{HH}=0.541$ \cite{donetti_hole_2011}). As a consequence the majority population in the QD ground state is now HH-like in all cases (no strain: HH=62\%/LH=32\%/SO=6\%, BC1: HH=70\%/LH=22\%/SO=8\%, and BC2: HH=73\%/LH=19\%/SO=8\%). 

The experimental metrics we use as a reference to assess our simulation results were extracted from the second dot (Q2) of the device in \cite{camenzind_hole_2022}. The experimental dot length was calculated using $l_{dot}=\hbar/\sqrt{m^{*}\Delta}$ (with $m^{*}$ = 0.45) and is in good agreement with the computed ones. Looking first at the energy splitting $\Delta$, we find that the inclusion of strain in the simulation domain is indispensable if one is to reproduce experimental measurements. The energy spacing for the unstrained qubit (0.8 meV) is indeed far below the values obtained for BC1 and BC2 (2.6 meV and 2.7 meV, respectively), the latter being only a few tenths of meV below the measured one (3.3 meV). 
Next, inspection of the $g$-factor values reveals that both strain hypotheses are qualitatively and quantitatively in relatively good agreement with measurements, while without strain the simulated and experimental values display large discrepancies. In fact, for BC1 and BC2, the strong structural confinement originating from the triangular-shaped cross-section effectively translates in a dominant $g_y$ and $g_z$ pair and a weaker $g_x$ term, in accordance with experimental observations. Moreover, from our computation we can also confirm that the principal axes for the $g$-factors remain aligned along the original $X$, $Y$, and $Z$ axes although they are slightly tilted in the $xy$-plane ($\approx20^{\circ}$), a feature that could be witnessed in experiments as well ($\approx12^{\circ}$). 

Regarding the RF, we report the value obtained when a transverse magnetic field along the $y$ axis and with a magnitude of 100 mT is applied to the qubit. Comparison between the experimental and simulated RF results suggests that the proposed compressive scenario (BC1) is the closest to reality. The calculated $f_{R}$ under this assumption (8.5 MHz) is indeed in a close range with the experimentally measured Rabi frequency (10 MHz), whereas, under BC2, the difference is much larger (104 MHz instead of 10 MHz). 

Eventually, we aim to reach the closest possible match between experiments and simulation. This remains a challenging task because, both on the experimental and modeling sides, numerous assumptions and calibrations must be made to deliver solid estimates. The simulation framework proposed in this paper rigorously describes the hole spin physics by using a 6-band instead of 4-band Luttinger-Kohn $k\cdot p$ Hamiltonian, which includes the SO bands (see Fig. \ref{fig:band_mixing}). In addition, the discretization step adopted in this work along with the localized description of the magnetic and strain fields guarantee that the ground state of the hole qubit and its figures of merit are accurately calculated. It is nonetheless worth mentioning that the SP simulations do not include non-idealities such as interface roughness, surface reconstruction, or defects inside the Si channel and at its interface with the oxides. These effects might have a sizable influence on the potential landscape of the quantum device, thus impacting the device metrics \cite{martinez_variability_2022}. Accounting for them will be the subject of a future study. For the device metrics of interest in this paper and under BC1, we could reach results relatively close experimental data. Yet more efforts are still required to get quantitatively closer to measurements. We recall that the simulated device differentiates itself from the experimental one which was designed to host two coupled quantum dots. In the DQD FinFET an additional gate (barrier gate) is added between the two plunger gates that localize the dots. The barrier gate is then used to fine-tune the magnitude of the exchange between the two coupled qubits along the transport direction $x$. The presence of this additional gate induces an additional electrostatic confinement inside the channel that is not considered in our simulations, but could increase both the energy spacing $\Delta$ and the $g$-factor $g_x$. Besides, the proximity of a second dot is expected to affect the overall shape and symmetry of the original QD. This might reorganize the spatial distribution of the basis states, hence modifying the $g$-factors and the RF map of the single dot. Furthermore, the sharp orthogonal edges and perfect interfaces used to define the oxide and metallic gates boundaries, partly due to the finite difference discretization of the domain, represent a simplification of the original device configuration. Simulation of the real DQD FinFET device under a realistic thermal deformation scenario (possibly compressive) with exactly scaled cross-section and remaining electrodes would be necessary to recover the exact strain distribution in the Si and hence to accurately characterize the quantum device and further improve the computed device metrics.

 To conclude our results we investigate the effect of different crystallographic orientations on the FinFET device metrics. In the experimental device, the $[110]$ crystal axis is aligned with the device transport direction ($x$ axis) so that the two remaining axes, $y$ and $z$, are aligned with the $[1\Bar{1}0]$ and $[001]$ directions, respectively. Besides the experimental alignment, labeled $[110]_x$, we define two additional configurations $[100]_x$ ($x=[100]$, $y=[011]$, and $z=[0\Bar{1}1]$), and $[101]_x$ ($x=[101]$, $y=[010]$, and $z=[0\Bar{1}1]$) and then compute their figures of merit. Based on our previous observation, the BC1 scenario is the most likely one, we then show in Fig. \ref{fig:comparison} principal g-factor values and Rabi frequency maps under that assumption for the novel crystal orientations. We first notice that the principal magnetic axes (dotted arrows) no longer align with the device axes (solid arrows). This phenomenon is expected because of the combined anisotropy of the Young's modulus, effective masses, and device geometry. The computed Rabi maps provide an overview of the interplay between the crystallographic and magnetic orientation that can be chosen to enhance the qubit performance and/or to comply with fabrication constraints such as restrictions in the magnetic field angle. For example, the $[101]_x$ configuration leads to a wider range of maximum $f_R$, while the $[100]_x$ one produces high $f_R$ only in the qubit transverse plane ($xy$-plane). We would like to emphasize that our modeling platform is equally suited to planar, FinFET, and nanowire-like quantum devices and thus can be leveraged to unravel the rich physics at play and the sweet-spots in hole spin qubit architectures within affordable computational time.

\section{\label{sec:Conclusion}Conclusion}
The linear thermal elastic simulations carried out in this work showed evidence that the channel of Si quantum devices capable of hosting hole qubits experiences non-negligible strain, although its exact nature remains to be determined. The presence of strain has sizable effect on the QD metrics and cannot therefore be omitted. The choice of the boundary conditions in the elastic model strongly influences the thermally induced stress within the FinFET. It moves the deformation maximum closer or further away from the center of the Si channel where the QD is formed. The calculation of the strain tensor from these displacements highlights the compressive character of the strain when only the substrate is kept at fixed position. It changes to tensile strain when the outer-faces of the metallic gates are fixed as well. 

Our theoretical study also outlined the dependence of the QD band mixing on the electrostatic landscape. This was achieved by considering three device geometries with different aspect ratios, but the same area, to change the confining potential in which the dot is created. At cryogenic temperature, we found that the thermally induced strain is more important than the QD geometry, having a strong impact on the band mixing of the ground state. Simulations under homogeneous static magnetic fields revealed that the qubit spin mixing depends on the magnetic orientation, which leads to different renormalizations of the spin-orbit basis states. Without any strain, the strong structural confinement originating from the triangular-shaped channel enhances the in-plane confinement, thus increasing the $g_{y}$ and $g_{z}$ factors in the cross section plane. The inclusion of strain causes a widening of the energy spacing between the ground and excited states, hence resulting in a more localized dot and enhanced $g_{x}$. The calculated Rabi frequency maps uncovered the rich physics arising from the varying electrostatic confinement due to the triangular channel and the fluctuating strain gradient experienced by the qubit.

Due to the absence of specific experimental data, it remains difficult to precisely estimate the strain behavior inside the Si channel of the quantum device. Besides the cooling-induced forces, residual strain originating from the oxide and/or metal deposition could be present, further complicating the thermal contraction model. Based on the obtained results, tensile strain in the Si channel appears to be advantageous since it leads to a substantial increase of the Rabi frequencies. This underlines the potential of applying strain engineering to such structures to induce an expansion of the semiconductor channel and positively affect the qubits performance. However, before drawing definitive conclusions, dephasing times due, for example, to charge noise should be thoroughly investigated to determine the hole spin qubit sweet spots.

\begin{acknowledgments}
We would like to thank Filip Mrcarica for the first implementation of the Pikus-Bir Hamiltonian and Dr. Youseung Lee for insightful discussions. This work was supported by the Swiss National Science Foundation under the NCCR SPIN (grant $\mathrm{n^\circ}$ 225153) and the QuaTrEx project (grant $\mathrm{n^\circ}$ 209358). The usage of CSCS computing resources under project s1119 is acknowledged.
\end{acknowledgments}

The data that support the findings of this article are openly available \cite{Bouquet}.

\appendix
\section{\label{AppendixA}Inclusion of magnetic field in the 6$\times$6 k$\cdot$p Hamiltonian}
We derive her the components of the wave vector \textbf{k} arising from the addition of a vector potential \textbf{A} to the Hamiltonian matrix describing the device of interest (Section \ref{subsec:General simulation}):

\begin{equation}
\label{eq:A1}
\boldsymbol{{\tilde{k}}}=\boldsymbol{k}+\frac{e}{\hbar}\textbf{A}.
\end{equation}

\subsection{Analytical expressions from the Peierls substitution}

With the magnetic field $B$=($B_{x}$,$B_{y}$,$B_{z}$), the gauge: $\textbf{A}=-(B_{z}y,0,B_{y}x−B_{x}y)$, and $\textbf{B} =\boldsymbol{\nabla}\times\textbf{A}$ we obtain for the different components of \textbf{k}

\begin{equation}
\label{eq:A2}
\tilde{k}_{x}=k_{x}-\frac{e}{\hbar}B_{z}y
\end{equation}
\begin{equation}
\label{eq:A3}
\tilde{k}_{y}=k_{y}
\end{equation}
\begin{equation}
\label{eq:A4}
\tilde{k}_{z}=k_{z}-\frac{e}{\hbar}(B_{y}x-B_{x}y).
\end{equation}

We then rewrite the $k_{i}k_{j}$ terms present in the Hamiltonian as $\frac{1}{2}(k_{i}k_{j}+k_{j}k_{i})$ and set $k_{i}j \propto \partial_i(j)=0$ for $i,j=x,y,z$ and $i\neq j$. If we only retain the linear terms in $B$, the contribution of \textbf{A} to the Hamiltonian becomes

\begin{equation}
\label{eq:A5}
\tilde{k}_{x}^{2}=k_{x}^{2}-\frac{2e}{\hbar}B_{z}yk_{x}\xrightarrow{\propto B}-\frac{2e}{\hbar}B_{z}yk_{x}
\end{equation}
\begin{equation}
\label{eq:A6}
\tilde{k}_{y}^{2}=k_{y}^{2}\xrightarrow{\propto B}0
\end{equation}
\begin{equation}
\label{eq:A7}
\tilde{k}_{z}^{2}=k_{z}^{2}-\frac{2e}{\hbar}B_{y}xk_{z}+\frac{2e}{\hbar}B_{x}yk_{z}\xrightarrow{\propto B}-\frac{2e}{\hbar}(B_{x}yk_{z}-B_{y}xk_{z})
\end{equation}

From these equations, expressions for the cross-terms can be found

\begin{equation}
\label{eq:A8}
\tilde{k}_{x}\tilde{k}_{y}+\tilde{k}_{y}\tilde{k}_{x}\xrightarrow{}-\frac{2e}{\hbar}B_{z}yk_{y}-\frac{e}{\hbar}B_{z}k_{y}y
\end{equation}
\begin{equation}
\label{eq:A9}
\tilde{k}_{y}\tilde{k}_{z}+\tilde{k}_{z}\tilde{k}_{y}\xrightarrow{}-\frac{2e}{\hbar}B_{y}xk_{y}+\frac{2e}{\hbar}B_{x}yk_{y}-
\frac{e}{\hbar}B_{x}k_{y}y
\end{equation}
\begin{multline}
\label{eq:A10}
\tilde{k}_{x}\tilde{k}_{z}+\tilde{k}_{z}\tilde{k}_{x}\xrightarrow{}\\
-\frac{e}{\hbar}B_{y}k_{x}x-\frac{2e}{\hbar}B_{y}xk_{x}+\frac{2e}{\hbar}B_{x}yk_{x}-\frac{2e}{\hbar}B_{z}yk_{z}
\end{multline}

The $k_{i}i$ ($\propto \partial_i(i)$) terms lead to constant terms and form the angular momentum operator \textbf{L} that enters Eq. (\ref{eq:two}) of the magnetic k$\cdot$p Hamiltonian $H_z$ \cite{luttinger_quantum_1956}.

\subsection{Discretized form of the vector potential contributions}

We perform the two-step substitution: $$k_{\alpha}\rightarrow-i\partial_{\alpha}$$ for $\alpha=x,y,z$ and $$\beta\partial_{\alpha}\Psi = \frac{\Psi_{i+1}-\Psi_{i-1}}{2\Delta\alpha}\beta_{i} $$ where $\Psi$ is the wave function vector.

Accounting now only for the non-diagonal terms and setting $T_{1}=e/\hbar$ to we have

\begin{equation}
\label{eq:A11}
\tilde{k}_{x}^{2}: i\frac{T_{1}B_{z}y_{ijk}}{\Delta x}(\Psi_{i+1jk}-\Psi_{i-1jk}) \propto L,M 
\end{equation}
\begin{equation}
\label{eq:A12}
\tilde{k}_{z}^{2}: i\frac{T_{1}}{\Delta z}(B_{y}x_{ijk}-B_{x}y_{ijk})(\Psi_{ijk+1}-\Psi_{ijk-1}) \propto L,M 
\end{equation}
\begin{equation}
\label{eq:A13}
\tilde{k}_{x}\tilde{k}_{y}+\tilde{k}_{y}\tilde{k}_{x}: i\frac{T_{1}B_{z}y_{ijk}}{\Delta y}(\Psi_{ij+1k}-\Psi_{ij-1k}) \propto N 
\end{equation}
\begin{equation}
\label{eq:A14}
\tilde{k}_{y}\tilde{k}_{z}+\tilde{k}_{z}\tilde{k}_{y}: i\frac{T_{1}}{\Delta y}(B_{y}x_{ijk}-B_{x}y_{ijk})(\Psi_{ij+1k}-\Psi_{ij-1k}) \propto N 
\end{equation}
\begin{multline}
\label{eq:A15}
\tilde{k}_{x}\tilde{k}_{z}+\tilde{k}_{z}\tilde{k}_{x}: i\frac{T_{1}}{\Delta x}(B_{y}x_{ijk}-B_{x}y_{ijk})(\Psi_{i+1jk}-\Psi_{i-1jk})\\+i\frac{T_{1}B_{z}y_{ijk}}{\Delta z} (\Psi_{ijk+1}-\Psi_{ijk-1}) \propto N 
\end{multline}

In the Dresselhaus-Kittel-Kip formulation (i.e., with basis vectors: $\{X\Uparrow,Y\Uparrow,Z\Uparrow,X\Downarrow,Y\Downarrow,Z\Downarrow\}$), this leads to 2 identical 3$\times$3 block diagonal Hamiltonians. 

\begin{multline}
\label{eq:A16}
\Psi_{ijk\pm1}=\pm i\frac{T_{1}}{\Delta z} \times\\
\begin{bmatrix}
L(B_{y}x-B_{x}y) & 0 & \frac{N}{2}B_{z}y\\
0 & L(B_{y}x-B_{x}y) & 0\\
\frac{N}{2}B_{z}y & 0 & M(B_{y}x-B_{x}y)
\end{bmatrix}
\end{multline}
\begin{multline}
\label{eq:A17}
\Psi_{ij\pm1k}=\pm i\frac{T_{1}}{\Delta y} \times\\
\begin{bmatrix}
0 & \frac{N}{2}B_{z}y & 0\\
\frac{N}{2}B_{z}y & 0 & \frac{N}{2}(B_{y}x-B_{x}y)\\
0 & \frac{N}{2}(B_{y}x-B_{x}y) & 0
\end{bmatrix}
\end{multline}
\begin{multline}
\label{eq:A18}
\Psi_{i\pm1jk}=\pm i\frac{T_{1}}{\Delta x} \times\\
\begin{bmatrix}
MB_{z}y & 0 & \frac{N}{2}(B_{y}x-B_{x}y)\\
0 & LB_{z}y & 0\\
\frac{N}{2}(B_{y}x-B_{x}y) & 0 & LB_{z}y
\end{bmatrix}
\end{multline}
where the $ijk$ index is omitted for $x$, $y$, and $z$. We further define $L=\frac{\hbar^2}{2m_{0}}(\gamma_{1}+4\gamma_{2})$, $M=\frac{\hbar^2}{2m_{0}}(\gamma_{1}-2\gamma_{2})$, and $N=6\gamma_{3}$.

\subsection{Calculation on the magnetic operator for the $g$-matrix creation}

Following the notation given in \cite{venitucci_electrical_2018}, we can calculate the matrix elements of $M_{1}$ using the finite difference method. For instance

\begin{multline}
\left\langle\psi_{1}|M_{1,y}|\psi_{2}\right\rangle= -\frac{1}{2\delta B}[\left\langle\psi_{1}|H(V,+\delta B\textbf{y})|\psi_{2}\right\rangle- \\
\left\langle\psi_{1}|H(V,-\delta B\textbf{y})|\psi_{2}\right\rangle]
\end{multline},
where the 6$\times$6 diagonal blocks of $H$ are given by Eq.~\ref{eq:two} and the off-diagonal ones by Eq.~\ref{eq:A16}-\ref{eq:A18} for the vector potential contribution to the qubit orbital functions.

\subsection{\label{Appendix4}Spin-orbit basis states 2D plots}

\begin{figure}[h]
\includegraphics[width=250pt]{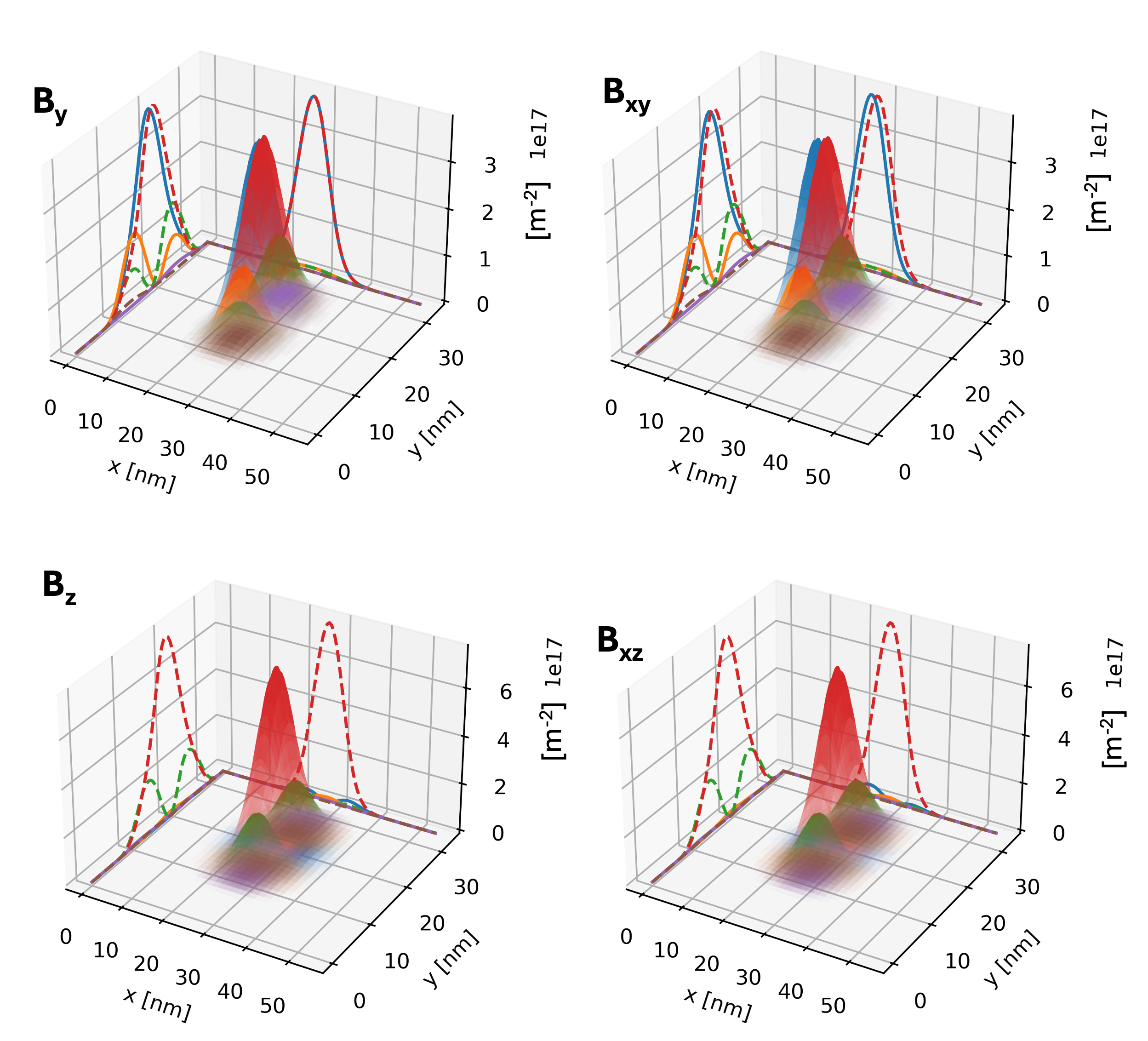}
\caption{\label{fig:app_spin_mixing}Representation of the qubit’s lowest energy state under static magnetic field ($B$ = 100 mT) oriented in the six independent magnetic field directions for Geo1. The central surface plots represent the qubit charge density in m$^{-2}$ average along the z axis. The 1D curves drawn at $x$ = 0 and $y$ = 42 are the charge densities along the dashed lines passing through the dot’s center, i.e, at $x$ = 27 nm and $y$ = 19 nm.}
\end{figure}

\begin{figure}[h]
\includegraphics[width=250pt]{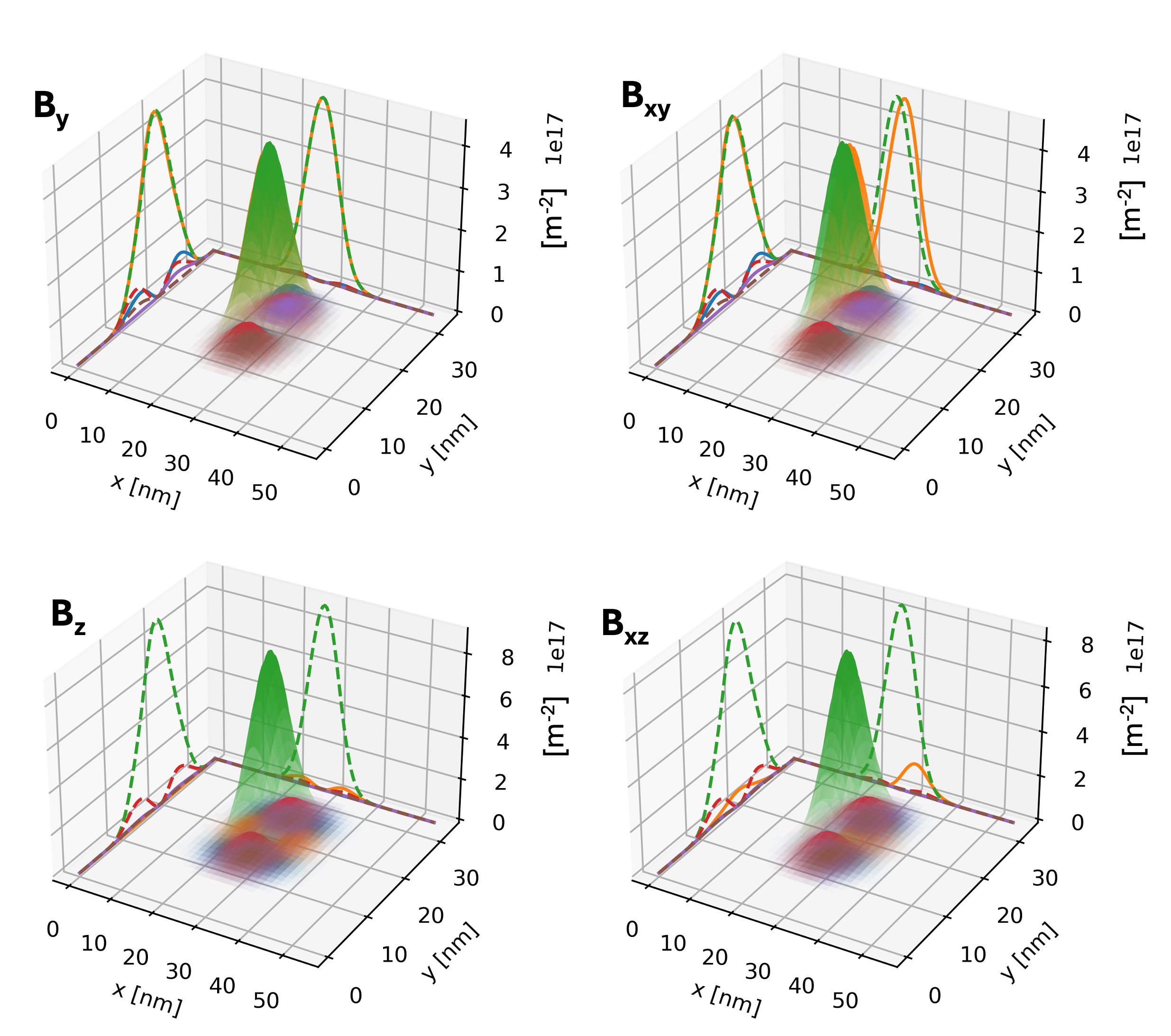}
\caption{\label{fig:app_spin_mixing_BC1}Same as Fig. \ref{fig:app_spin_mixing} for BC1.}
\end{figure}

\begin{figure}[h]
\includegraphics[width=250pt]{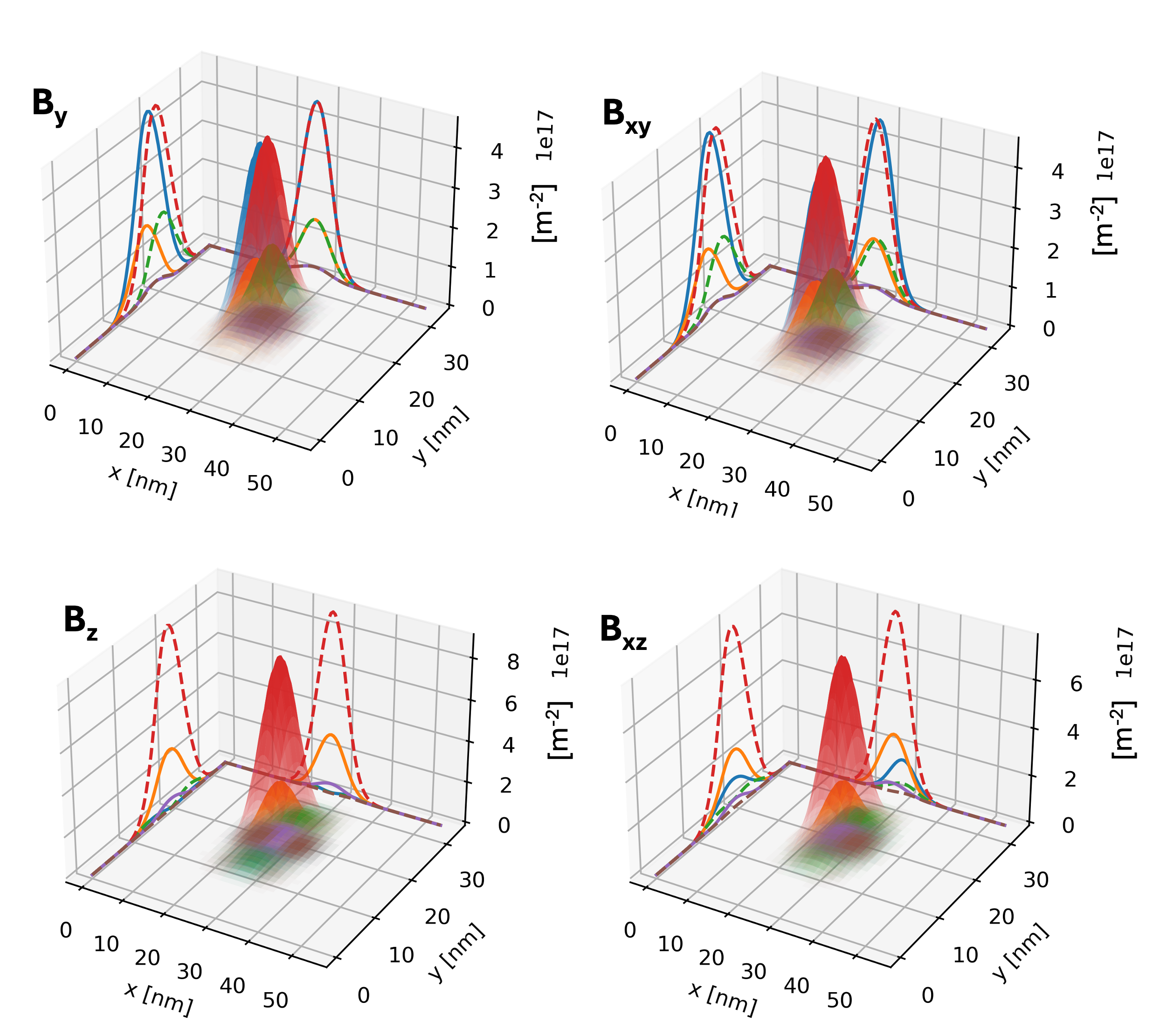}
\caption{\label{fig:app_spin_mixing_BC2}Same as Fig. \ref{fig:app_spin_mixing} for BC2.}
\end{figure}

In Fig. \ref{fig:app_spin_mixing}-\ref{fig:app_spin_mixing_BC2} we plot the qubit ground state spin-orbit basis states obtained for homogeneous magnetic fields oriented along $y$, $z$, $xy$, and $xz$. The qubit g-tensor can then be reconstructed from the energy spacing obtained for each case (see \ref{subsec:General simulation}).

\subsection{\label{Appendix5}Strain tensor maps}

\begin{figure}
\includegraphics[width=250pt]{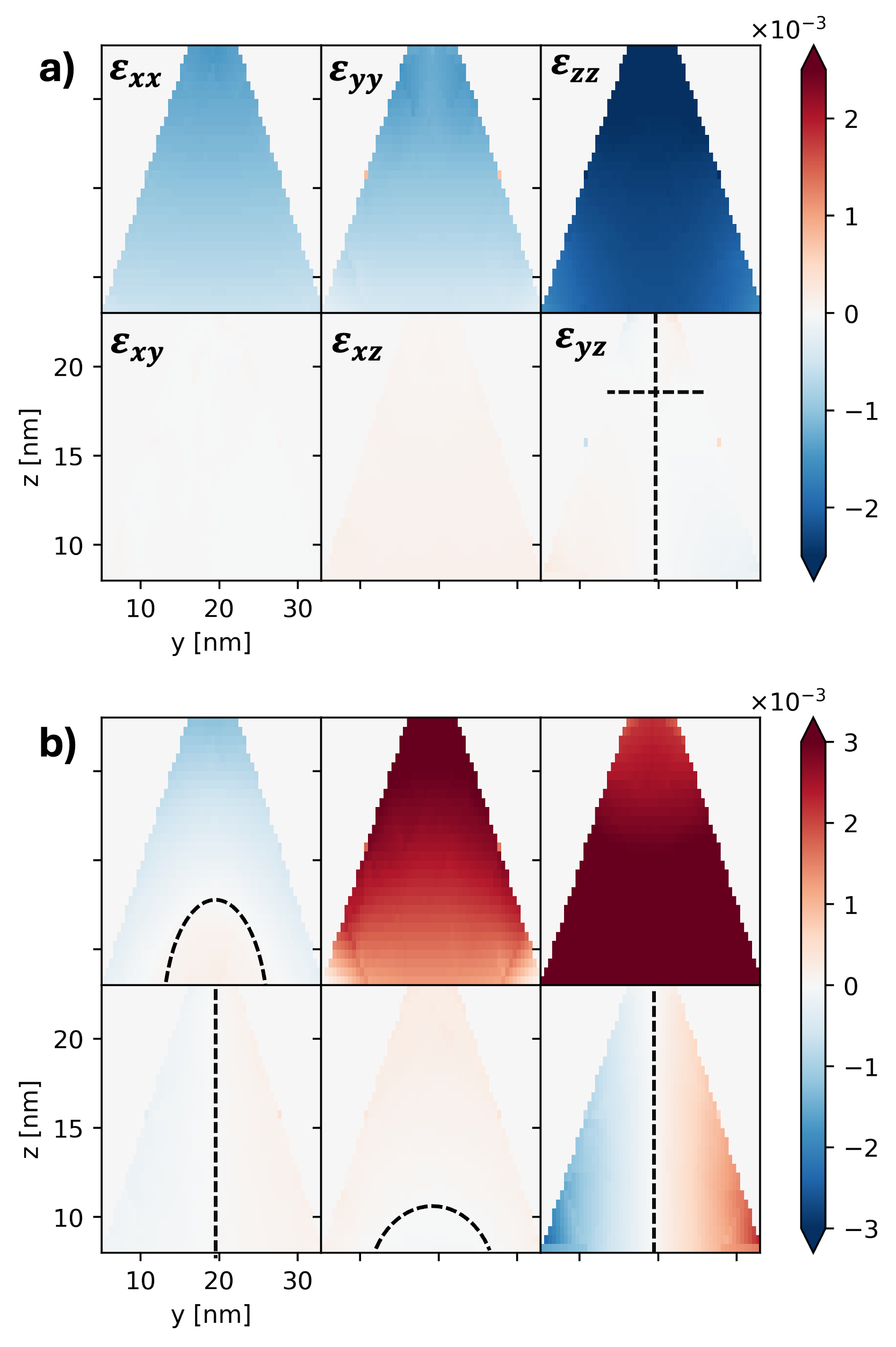}
\caption{\label{fig:strain_22} Maps of the strain tensor components along the device cross-section at $x$ = 22 nm for BC1 a) and BC2 b). The dotted black lines depict the demarcation between positive (expansion) and negative (compression) strain.}
\end{figure}

\begin{figure}
\includegraphics[width=250pt]{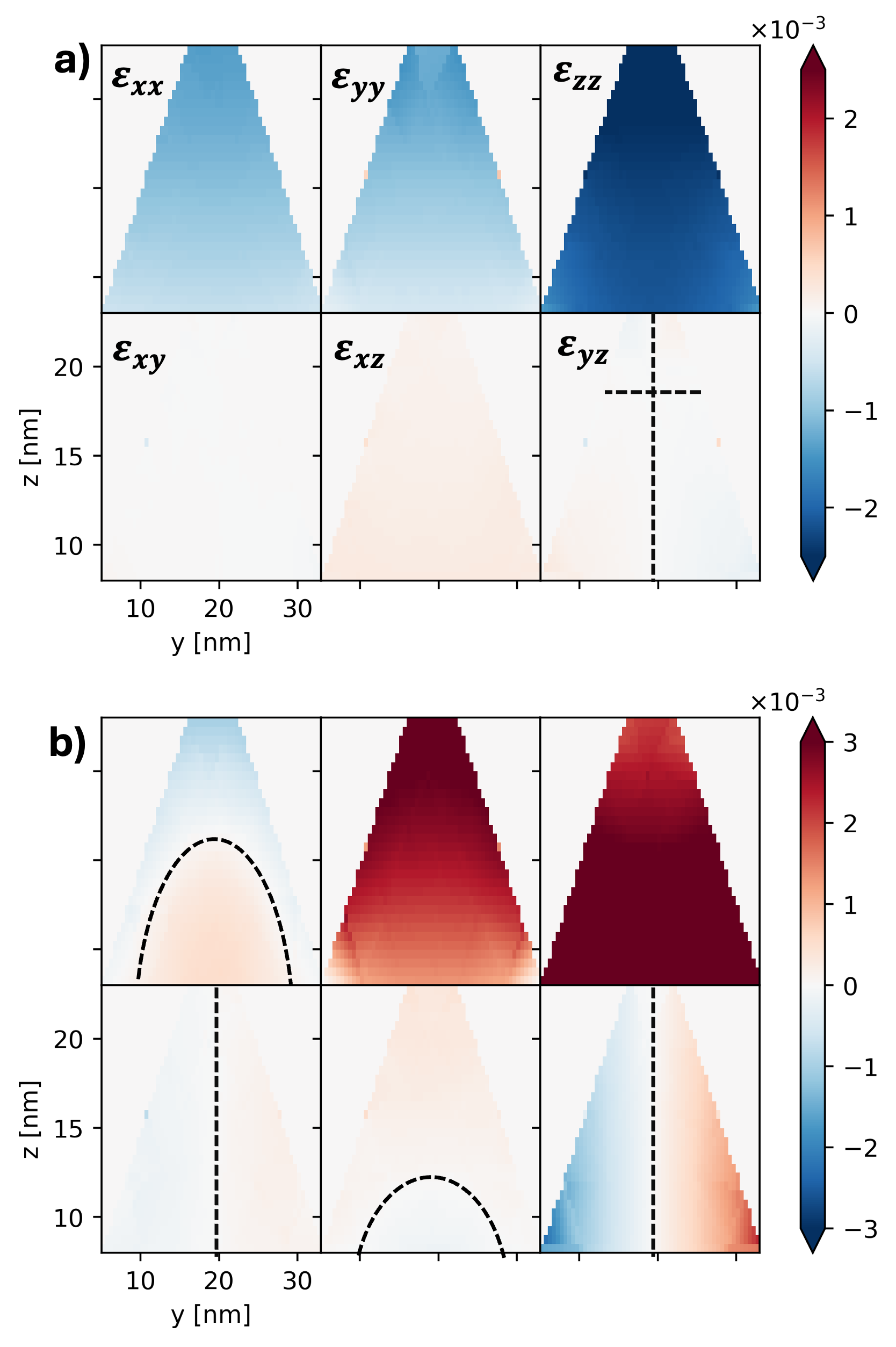}
\caption{\label{fig:strain_18} Maps of the strain tensor components along the device cross-section at $x$ = 18 nm for BC1 a) and BC2 b). The dotted black lines depict the demarcation between positive (expansion) and negative (compression) strain.}
\end{figure}

In Fig. \ref{fig:strain_22} and Fig. \ref{fig:strain_18}, we show the plots of the strain tensor along the cross-sections situated at $x$ = 22 nm and $x$ = 18 nm, respectively. Along the high symmetry plane at $x$ = 27 nm, the mixed components $\varepsilon_{xy}$, $\varepsilon_{xz}$ and $\varepsilon_{yz}$ are close or equal to zero, whereas away from the middle a combination of compressive and tensile strain can be observed for both strain scenarios in the qubit plane.

\clearpage

\bibliography{bibliography.bib}
%\printbibliography
\end{document}